\documentclass[a4paper,fleqn,usenatbib]{mnras}
\usepackage{newtxtext,newtxmath}
\usepackage[T1]{fontenc}
\usepackage{ae,aecompl}

\usepackage{epsfig,amsmath,amsfonts,amssymb,graphicx,subfigure,enumitem,url, deluxetable}
\usepackage{epstopdf}
\usepackage{hyperref}
\usepackage{deluxetable}

\title[Study of three eclipsing polars]{Optical and X-ray studies of three  polars:\\ RX J0859.1+0537, RX J0749.1-0549, and RX J0649.8-0737}
\author[Joshi et al.]{
Arti Joshi$^{1}$\thanks{E-mail: arti@aries.res.in},
J. C. Pandey$^{1}$,
Ashish Raj$^{2}$,
K. P. Singh$^{3}$,
G. C. Anupama$^{2}$,  
\newauthor ~and H. P. Singh$^{4}$
\\~\\
$^{1}$Aryabhatta Research Institute of observational sciencES, Manora Peak, Nainital 263 001, India\\
$^{2}$Indian Institute of Astrophysics, Koramangala, Bangalore 560 034, India\\
$^{3}$Indian Institute of Science Education and Research, Mohali, India\\
$^{4}$Department of Physics and Astrophysics, University of Delhi, 110007 Delhi, India
}

\date{Accepted XXX. Received YYY; in original form ZZZ}

\pubyear{2019}

\begin{document}
\label{firstpage}
\pagerange{\pageref{firstpage}--\pageref{lastpage}}
\maketitle


\begin{abstract}
	We present optical photometric and spectroscopic observations, and an analysis of archival X-ray data of three  polars: RX J0859.1+0537, RX J0749.1-0549, and RX J0649.8-0737. Optical light curves of these three polars reveal eclipse features that are deep, total, and variable in shape. The optical and X-ray modulations of RX J0859.1+0537, RX J0749.1-0549, and RX J0649.8-0737 are both found to occur at the orbital periods of 2.393$\pm$0.003 hrs, 3.672$\pm$0.001 hrs, and 4.347$\pm$0.001 hrs, respectively. RX J0859.1+0537 is found to be an eclipsing polar which lies in the region of the period gap, whereas RX J0749.1-0549 and RX J0649.8-0737 are found to be long-period eclipsing polars above the period gap. The eclipse length is found to be 61 min for RX J0749.1-0549 in the R-band, which is the highest among the long period eclipsing polars. The radius of the eclipsed light source is found to be more than the actual size of the white dwarf for these three systems, indicating that the eclipsed component is not only the white dwarf but also appears to include the presence of an extended accretion region.  Optical spectra of these systems show the presence of high ionization emission lines along with the strong Balmer emission lines with an inverted Balmer decrement. Cyclotron harmonics are also detected in the optical spectra from which we infer magnetic field strength of the surface of the white dwarf to be 49$\pm$2 MG, 43.5$\pm$1.4 MG, and 44$\pm$1 MG for RX J0859.1+0537, RX J0749.1-0549, and RX J0649.8-0737, respectively.

\end{abstract}

\begin{keywords}
accretion, accretion discs -- novae, cataclysmic variables -- binaries: eclipsing -- stars: individual: RX J0859.1+0537 -- stars: individual: RX J0749.1-0549 -- stars: individual: RX J0649.8-0737.
\end{keywords}


\section{Introduction}
\label{sec:intro}
 
Polars are a class of Magnetic Cataclysmic Variables (MCVs) in which material transfers from Roche lobe-filling red dwarf secondary to a highly magnetized white dwarf. These binary systems are synchronously locked and the formation of an accretion disc is prevented due to a high magnetic field ($>$ 10 MG) on the surface of the white dwarf (WD). The accreted matter follows the magnetic field lines of the WD when the magnetic pressure exceeds the ram pressure \citep[see][for a full review of MCVs]{Cropper90, Warner95}. The optical and near-infrared radiation is dominated by the cyclotron emission, while X-ray radiation is dominated by bremsstrahlung emission in hard X-rays. Both bremsstrahlung and cyclotron radiation originate from the  magnetically channeled accretion column near the WD in the post-shock region. Cyclotron radiation is highly circularly polarized, reaching 50\% in optical bands \citep[see][]{Cropper90}. The cyclotron cooling is extremely efficient \citep{Lamb79, Woelk92}, therefore, a part of the downward emitted hard X-rays and cyclotron radiation is absorbed by the WD surface and re-emitted in the soft X-rays or in the extreme UV. The soft X-ray photons are mostly emitted near the base of the post-shock flow and thus a majority of the polars are found to show a `soft X-ray excess'.

Observationally, various polars reveal a structured X-ray and photometric eclipse profile. These eclipse profiles are mostly noticed in the high-inclination systems where the white dwarf, the accretion region, and the accretion streams are occulted by the secondary star and provide information about the spatial structures and the binary parameters of the system. Light curves of the eclipse contain information about the structure and the brightness distribution along the stream. The spectroscopic observations of polars can also give further insights into the accretion region. The optical spectra of polars exhibit strong Balmer emission lines along with He I and He II emission lines. The majority of  polars have orbital periods shorter than the ``period gap (2-3 hrs)". 

The catalogue of \citet[][update RKcat7.24, 2016, RK]{Ritter03}\footnote{https://wwwmpa.mpa-garching.mpg.de/RKcat/cbcat} identified 119 confirmed and 31 uncertain polars. From this catalogue, we took three poorly studied polars namely RX J0859.1+0537 (hereafter RX J0859), RX J0749.1-0549 (hereafter RX J0749), and RX J0649.8-0737 (hereafter RX J0649) for a detailed study. These three systems are identified as polars on the basis of their hardness ratio from  {\textit ROSAT} observations \citep{Beuermann95,Motch98}, and until now there has been no detailed optical and X-ray study carried for these sources, and therefore their exact nature has not been confirmed. The poor knowledge of these sources encouraged us to do the optical (photometric and spectroscopic) and X-ray study. A summary of the available information on these sources is given below.  

\subsection{RX J0859.1+0537}\label{RX J0859}
RX J0859 was identified as a polar \citep{Beuermann95, Szkody05} using optical spectroscopy and hardness ratio in X-rays. It is located at a distance of 441$\pm$30 pc \citep{GaiaCollaborationLindegren18}. From SDSS spectra obtained during 2003, \citet{Szkody05} found the orbital period of 1.1 hrs but later \citet{Gansicke09} re-analysed the SDSS spectrum and derived the orbital period of $\sim$2.4 hrs. \citet{Harrison15} found that the WISE (Wide-field Infrared Survey Explorer) W1 (=3.4 $\mu$m wavelength band) light curve of RX J0859 folded reasonably well with a period of 1.1 hrs, while the source was not found to vary in WISE W2 (=4.6 $\mu$m wavelength band) light curve. \citet{Harrison15} have also estimated the field strength of RX J0859 between 20 -35 MG after attributing the humps in the W1 light curve to cyclotron emission.

\subsection{RX J0749.1-0549}\label{sec:RX J0749}
RX J0749 was first identified  by \citet{Motch98} among the ROSAT Galactic Plane Survey sources located at low galactic latitudes. They reported it to be a long period polar with an orbital period of $\sim$3.6 hrs.  Using Gaia DR2 parallax the distance of RX J0749 is calculated to be 1168$\pm$164 pc \citep{GaiaCollaborationLindegren18}.

\subsection{RX J0649.8-0737}\label{sec:RX J0649}
\citet{Motch98} characterized RX J0649 as a polar with  an orbital period of $\sim$ 4.4 hrs. It is located at a distance of  611$\pm$38 pc \citep{GaiaCollaborationLindegren18}.  WISE W1 light curves of RX J0649 show an ellipsoidal variation with a period of 4.4 hrs \citep[see][]{Harrison15}. However, the WISE W2 light curve, seems to be deeper, and has a shorter-lived minimum than in W1. \citet{Harrison15} also suggested that the WISE W2 light curve was more consistent with a strong cyclotron emission with magnetic field  B $\leq$ 26 MG.  From near-IR colors they reported that the secondary star is consistent with an early M-type star.

This paper is organized as follows: we summarize our optical observations and data reduction in the next section. Section 3 contains analyses and the results of the optical and X-ray data. Finally, we present discussion and conclusions in sections 4 and 5, respectively.

\begin{table}
\centering
\tiny
\setlength{\tabcolsep}{0.007in}
\caption{A log of photometric, spectroscopic, and X-ray observations of RX J0859.1+0537, RX J0749.1-0549, and RX J0649.8-0737.\label{tab:obslog}}
\begin{tabular}{lllllllll}
\hline\hline
  Object &~~~Date of &Telescope&Instrument&Filter/ &Integration      &    ~~~Time              &Time  \\
         &~~~~Obs. &         &          &band   &Time (s)         & ~~(HJD$_{t}$)          &span(hr)     \\
\hline\\
RX J0859                    & ~2014 Jan 30 & ~1.3-m  & CCD                & ~R 	        &~~250    &2456688.00 &~~2.5\\
		                        & ~2014 Jan 31 & ~1.3-m  & CCD                & ~R 	        &~~200    &2456689.00 &~~2.6\\
                            & ~2014 Feb 23 & ~1.3-m  & CCD                & ~R 	        &~~250    &2456712.00 &~~3.4\\ 
                            & ~2014 May 03 & ~1.04-m & CCD                & ~R 	        &~~300    &2456781.00 &~~1.8\\
                            & ~2015 Mar 27 & ~1.04-m & CCD                & ~R 	        &~~300    &2457105.00 &~~3.7\\                                
                            & ~2015 Apr 09 & ~1.04-m & CCD                & ~R 	        &~~300    &2457122.00 &~~3.3\\
                            & ~2015 Apr 10 & ~1.04-m & CCD                & ~R 	        &~~300    &2457123.00 &~~2.8\\
                            & ~2017 Dec 22 & ~2.01-m & HFOSC/Gr07         &~380-780 nm  &~~3600   &2458110.43 &~~1.0\\
			                      & ~1996 Apr 24 & ~ROSAT  & HRI                &~0.1-2.2 keV &~~32846  &2450197.95 &~~433.2\\                            
RX J0749                    & ~2014 Jan 30 & ~1.3-m  & CCD               	& ~R          &~~250    &2456688.00 &~~2.6\\ 
		                        & ~2014 Jan 31 & ~1.3-m  & CCD              	& ~R          &~~250    &2456689.00 &~~2.8\\
                            & ~2014 Apr 24 & ~1.3-m  & CCD                & ~R          &~~250    &2456772.00 &~~2.5\\  
	    	                    & ~2014 Apr 25 & ~1.3-m  & CCD              	& ~R          &~~250    &2456773.00 &~~2.4\\
                            & ~2015 Jan 23 & ~1.04-m & CCD             	  & ~R          &~~300    &2457046.00 &~~4.9\\ 
                            & ~2015 Apr 16 & ~1.04-m & CCD             	  & ~R          &~~300    &2457129.00 &~~1.8\\
                            & ~2017 Dec 22 & ~2.01-m & HFOSC/Gr07         &~380-780 nm  &~~3600   &2458110.34 &~~1.0\\
			                      & ~1997 Apr 09 & ~ROSAT  & HRI                &~0.1-2.2 keV &~~45490  &2450548.02 &~~203.1\\
RX J0649                    & ~2015 Nov 06 & ~1.3-m  & CCD                & ~R          &~~200    &2457333.00 &~~3.7\\
		                        & ~2015 Nov 21 & ~1.3-m  & CCD                & ~R          &~~200    &2457348.00 &~~4.5\\
                            & ~2015 Dec 19 & ~1.3-m  & CCD                & ~R          &~~250    &2457376.00 &~~3.3\\ 
                            & ~2017 Dec 22 & ~2.01-m & HFOSC/Gr07         &~380-780 nm  &~~3600   &2458110.19 &~~1.0\\
			                      & ~1997 Sept 23 &~ROSAT  & HRI               &~0.1-2.2 keV &~~17080  &2450714.97 &~~4.7\\
 
\hline
\end{tabular}
\vskip.4cm 
{{\bf \it Note.-}} HJD$_{t}$ for HFOSC and HRI instrument is the start time of the observations, whereas for optical observation it is observation day.
\end{table}

\begin{table}
\centering
\small
\setlength{\tabcolsep}{0.2in}
\caption{Comparison stars used for the differential photometry for RX J0859.1+0537, RX J0749.1-0549, and RX J0649.8-0737.\label{tab:obj_comp}}
\begin{tabular}{lllllllll}
\hline\hline
Object & Reference  &~B    &~R \\
       & USNO-B1.0  &(mag)&(mag) \\
\hline\\
RX J0859                    & 0956-0174607 &  19.42  &  16.93  \\
`C1'		                    & 0956-0174599 &  17.55  &  16.34  \\
`C2'                        & 0956-0174609 &  16.83  &  16.13  \\ 
RX J0749                    & 0841-0159780 &  17.36  &  17.75  \\  
`C1'	    	                & 0841-0159756 &  17.30  &  16.96  \\
`C2'                        & 0841-0159759 &  16.46  &  16.17  \\ 
RX J0649                    & 0823-0138560 &  18.80  &  16.69  \\
`C1'		                    & 0823-0138520 &  17.58  &  16.31  \\
`C2'                        & 0823-0138559 &  17.02  &  16.51  \\ 
\hline
\end{tabular}
\vskip.4cm 
{{\bf \it Note.-}} Here, C1 and C2 stands for comparison 1 and 2 for each program star.
\end{table}

\begin{table}
\centering
\small
\setlength{\tabcolsep}{0.05in}
\caption{The times of eclipse minima of RX J0859.1+0537, RX J0749.1-0549, and RX J0649.8-0737.\label{tab:eclipsetime_07_08}}
\begin{tabular}{lllllllll}\hline\hline
   Object &      ~~Date  &         Eclipse Minima &   Cycles\\
          &      of Obs. &         ~~~~(HJD$_{t}$+)   &     \\
  \hline\\
RX J0859       &2014 Jan 30   & 0.426723$\pm$0.044333       &~~0           \\    
	              &2014 Jan 31   & 0.426093$\pm$0.041722      &10.00           \\ 
                &2014 Feb 23   & 0.318837$\pm$0.028185      &239.00       \\
	              &2015 Mar 27   & 0.131961$\pm$0.022238      &4169.97     \\
	              &2015 Mar 27   & 0.231166$\pm$0.042367      &4170.96      \\
	              &2015 Apr 09   & 0.124104$\pm$0.019132      &4340.01       \\
	              &2015 Apr 10   & 0.221810$\pm$0.047712      &4350.99        \\
RX J0749	      &2014 Jan 31   & 0.285211$\pm$0.002894      &~~0           \\
		            &2014 Apr 24   & 0.107516$\pm$0.002894      &537.02      \\
	 	            &2015 Jan 23   & 0.308878$\pm$0.003472      &2314.95     \\
	 	            &2015 Apr 16   & 0.126693$\pm$0.003472      &2851.95     \\
RX J0649        &2015 Nov 06   & 0.451228$\pm$0.067288      &~~0           \\
	              &2015 Nov 21   & 0.479756$\pm$0.051732      &81.97  \\
                &2015 Dec 19   & 0.329473$\pm$0.079010      &233.88  \\ 
\hline
\end{tabular}
\vskip.4cm 
  {{\bf \it Note.-}} HJD$_{t}$ is similar to the values defined in the Table \ref{tab:obslog}.
\end{table}


\section{Observations And Data Reduction}

\subsection{Optical Photometry} 

R-band photometric observations of these sources were obtained in 2014 and 2015 using the 1.04-m Sampuranand Telescope (ST) at ARIES, Nainital  \citep[][]{Sinvhal75} and 1.3-m Devasthal Fast Optical Telescope (DFOT) located at  Devasthal, Nainital, India \citep{Sagar11}. A detailed log for photometric observations is given in Table \ref{tab:obslog}. The ST has Ritchey-Chretien (RC) optics with a f/13 beam at the Cassegrain focus. The ST is equipped with a 2k$\times$2k Andor CCD (read noise=13.7 $e^{\rm -}$ $pixel^{\rm -1}$ and gain=10 $e^{\rm -}$/Analog to Digital Unit (ADU)). CCD covers a field-of-view (fov) of 13\hbox {$^\prime$}$\times$13\hbox {$^\prime$} and its each pixel has a dimension of 24 $\mu$m$^2$. Observations were carried out in a 2$\times$2 pixel$^2$ binning mode in order to increase the signal-to-noise ratio. 
The 1.3-m DFOT is also RC design with f/4 beam at the Cassegrain focus. The DFOT is equipped with a 2k$\times$2k Andor CCD (read noise= 7$e^{\rm -}$ $pixel^{\rm -}$  and gain=2 $e^{\rm -}$/ADU at 1 MHz readout speed). Each pixel size of the CCD is 13.5 $\mu$m$^2$ and covers a total field of $\sim$18\hbox {$^\prime$}$\times$18\hbox {$^\prime$}. Photometric observations of RX J0649 were also carried out from DFOT with a 512 $\times$ 512 Andor CCD (read noise = 6.1 $e^{\rm -}$ $pixel^{\rm -}$ and gain=10 $e^{\rm -}$/ADU at 1 MHz readout speed) with a pixel size of 13.5 $\mu$m$^2$ and the fov of the CCD is 5 arcmin on the sky. Several bias and twilight sky flat frames were taken during the observing runs. The pre-processing (bias subtraction, flat fielding, and cosmic ray removal) of the raw photometric data were performed using IRAF \footnote{IRAF is distributed by the national optical astronomy observatories, USA.} software. 
 Differential photometry (variable minus comparison star) was performed by using a comparison star in the same field. The details of the adopted comparison stars for each source are given in Table \ref{tab:obj_comp}. The instrumental magnitudes of the target and two comparison stars labelled as `C1' and `C2' were extracted from the images. Then the R-band magnitudes of observations were computed with respect to comparison star `C1'. The nightly variations ($\sigma$) of `C1-C2' were found to be in the range of 0.005-0.01 (for RX J0859), 0.007-0.01 (for RX J0749), and 0.009-0.01 (for RX J0649), respectively. The standard magnitude of `C1' and `C2' were taken from USNO-B1.0 catalogue (see Table \ref{tab:obj_comp}). The constant brightness of comparison stars was checked by inspecting the light curves of `C1-C2' for each source separately. 

\begin{figure*}
\centering
\subfigure[]{\includegraphics[width=58mm]{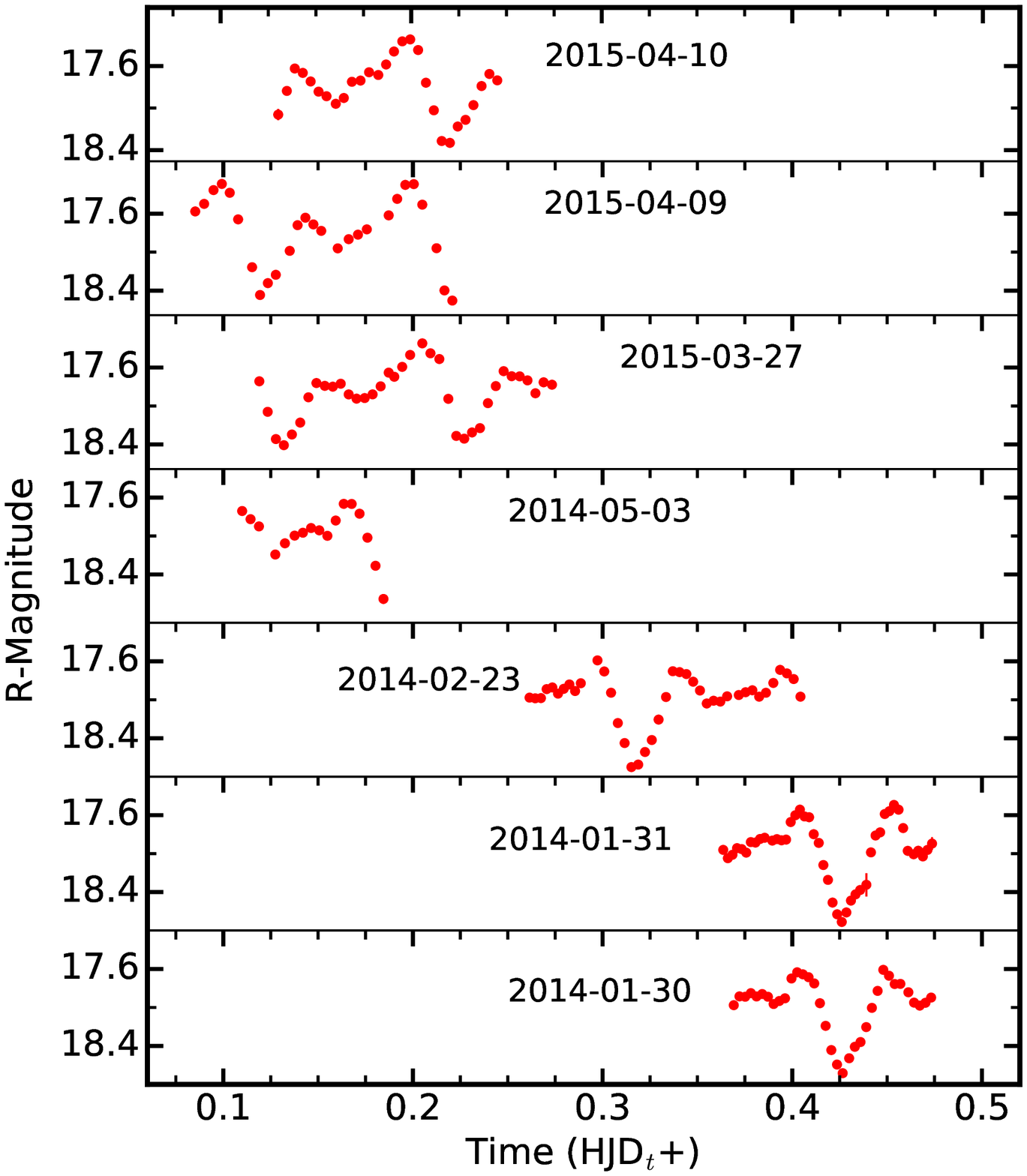}\label{fig:optlc_rxj08}}
\subfigure[]{\includegraphics[height=60mm, width=58mm]{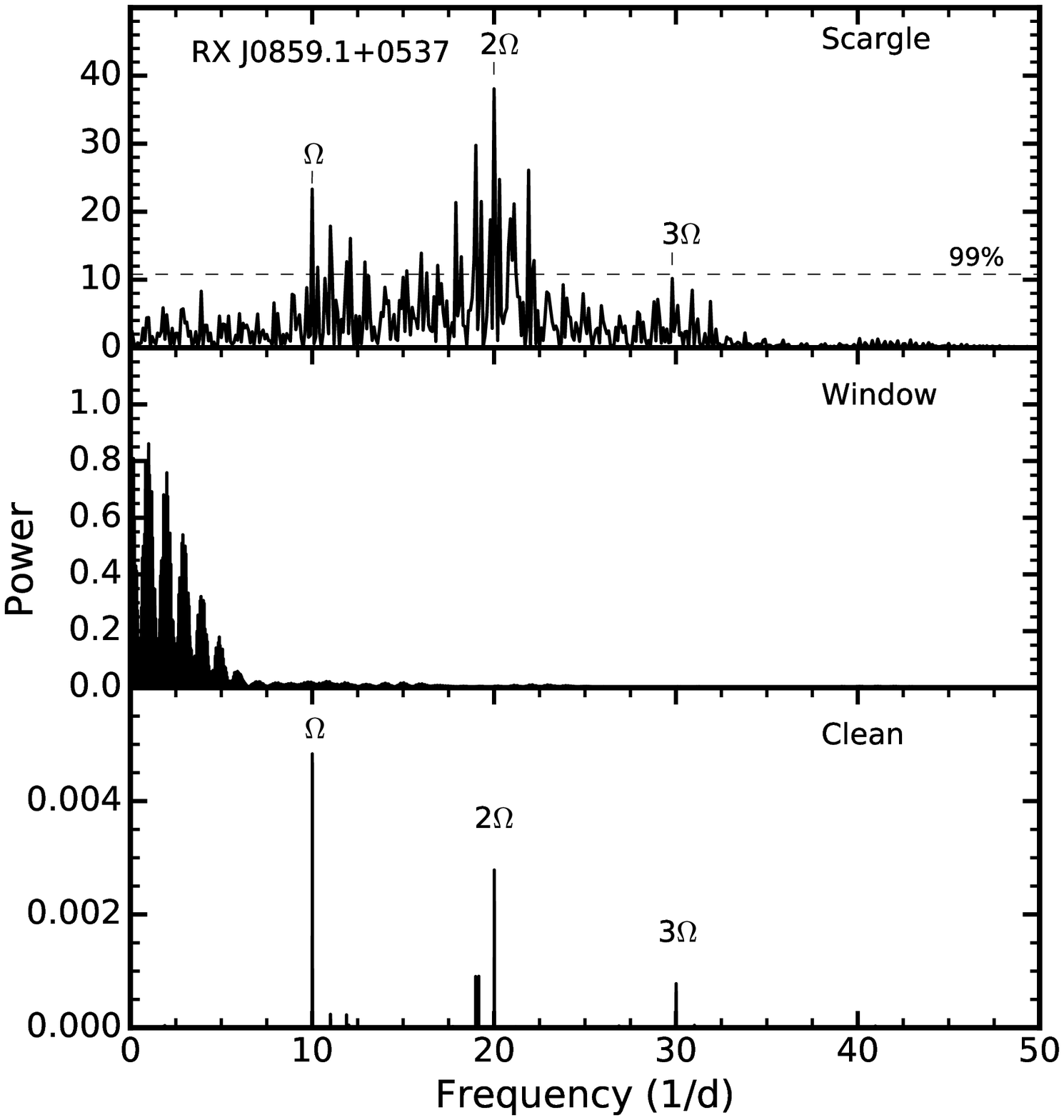}\label{fig:optps_rxj08}}
\subfigure[]{\includegraphics[height=60mm,width=58mm]{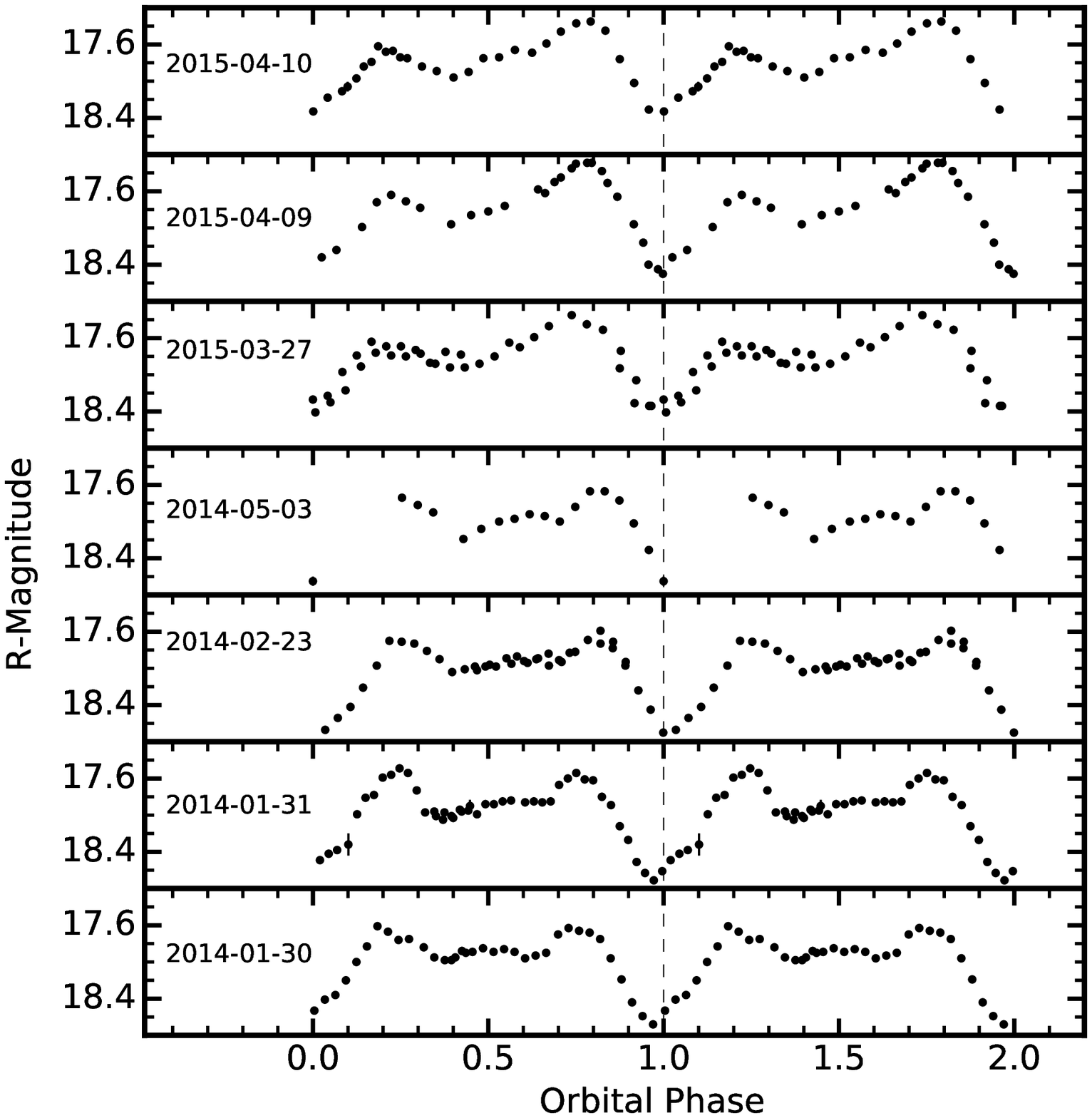}\label{fig:optfoldlc_rxj08}}
\caption{(a) R-band light curve of RX J0859.1+0537, (b) Lomb-Scargle power spectra (top panel), window function (middle panel), and CLEANed power spectra (bottom
  panel) of RX J0859.1+0537, and (c) folded R-band light curve of RX J0859.1+0537. The date of observations are mentioned near each light curve. The HJD$_{t}$ of each observation  is given in Table \ref{tab:obslog}.} 
\label{fig:optlc}
\end{figure*}

\subsection{Optical Spectroscopy} 

Long-slit low-resolution optical spectra for these polars were obtained on 22 December 2017 with the 2.01-m Himalayan Chandra Telescope at IAO, Hanle which is equipped with the Hanle Faint Object Spectrograph and Camera (HFOSC).  We used the Gr7 (3800-7800 \AA) grism  with resolution of 1330 for our observations. Exposure time was 3600 s for each observation in order to obtain good S/N spectra. Spectrophotometric standards were also observed during our observations. Detailed information about the spectroscopic observations is provided in Table \ref{tab:obslog}. The spectra of target and standard stars were extracted in the standard manner using IRAF tasks. In the extracted one dimensional spectrum the abscissa corresponds to the pixel number and ordinate is in the form of intensity. Lamp spectra of FeAr have been used for the wavelength calibration. The wavelength calibration was accomplished by using the IRAF-task {\sc identify}, which determines the pixel-to-wavelength solution for arc spectrum.  To execute this solution to the extracted spectrum of the standard star and target, the task {\sc dispcor} is used. To check the wavelength calibration, the night sky emission lines 5577 \AA, 6300 \AA, and 6363 \AA ~were used  and an appropriate shift is applied where necessary.  The flux calibration was performed by using the tasks {\sc standard}, {\sc sensfunc}, and {\sc calibrate} in IRAF. Therefore, the flux calibrated one-dimensional spectra finally provide the distribution of flux over wavelength.  The telluric lines were not removed from the spectra.

\subsection{ X-ray Data}
The archival X-ray data were obtained from the {\textit ROSAT} observations of RX J0859, RX J0749, and RX J0649 that were carried out on 24 April 1996, 09 April 1997, and 23 September 1997, respectively using the High-Resolution Imager \citep[HRI;][]{Zombeck95}.
A log of observations is given in  Table \ref{tab:obslog}.  {\textit ROSAT} has $\sim$ 40$^\prime$ diameter fov with the HRI in the focal plane observing in the energy range of $\sim$ 0.1-2.2 keV. HRI consisted of two cascaded microchannel plates (MCPs) with a crossed grid position readout system and provided a 38$^\prime$$\times$38$^\prime$ field-of-view with $\sim$ 2$^\prime$$^\prime$  spatial resolution (FWHM). HRI used a crossed grid detector with a position accuracy to 25 micrometers. This instrument had also a very low or negligible energy resolution but provided time resolution down to 61 $\mu$s relative to the ROSAT spacecraft clock.  The X-ray image and the light curves of these sources were extracted by using the {\sc xselect} task in the {\sc ftools} package. The X-ray light curves of each source were obtained using a circular region with a radius of 25$^\prime$$^\prime$ centered  on the source. The background regions were taken from a source-free region around the source in the same data.  
Finally, the background subtracted X-ray light curves were obtained by using the task {\sc lcmath}. The temporal binning for each light curve was 16 s.



\begin{table}
\caption{Periods corresponding to dominant peaks in the power spectra of RX J0859.1+0537, RX J0749.1-0549, and RX J0649.8-0737 obtained from X-ray and optical data.\label{tab:ps}}
\scriptsize
\setlength{\tabcolsep}{0.02in}
\begin{tabular}{cccccccccccccc}
\hline\hline \\
Object&&Period&&\multicolumn{2}{c}{\textbf{SCARGLE}} &&& \multicolumn{2}{c} {\textbf{CLEAN}}\\
\cline{5-6} \cline{8-10}\\
      && (hr)      &&Photometry&ROSAT&&& Photometry&ROSAT\\
\hline\hline
RX J0859 &&$P$$_\Omega$ && 2.40$\pm$0.02 &2.43$\pm$0.04 &&& 2.398$\pm$0.020& 2.393$\pm$0.003\\
         &&$2P$$_\Omega$&& 1.20$\pm$0.01 &1.23$\pm$0.02 &&& 1.203$\pm$0.001& 1.261$\pm$0.001 \\
         &&$3P$$_\Omega$&& 0.80$\pm$0.01 &0.80$\pm$0.01 &&& 0.799$\pm$0.003& 0.812$\pm$0.003 \\
RX J0749 &&$P$$_\Omega$ && 3.75$\pm$0.11& 3.68$\pm$0.02 &&& 3.672$\pm$0.001& 3.675$\pm$0.002\\
RX J0649 &&$P$$_\Omega$ && 4.32$\pm$0.20 &4.34$\pm$0.13 &&& 4.310$\pm$0.002& 4.347$\pm$0.001\\
         &&$3P$$_\Omega$&& 1.42$\pm$0.14 &1.26$\pm$0.10 &&& 1.482$\pm$0.001& 1.251$\pm$0.001 \\
\hline
\end{tabular}
\end{table}

\section{Analysis and Results}

\subsection{RX J0859.1+0537}
\label{sec:Analysis_RXJ0859}
\subsubsection{Optical Photometry}
\label{sec:phot08}
We obtained a total of 20.1 hrs of photometric data from our observations carried out over seven nights.  R-band photometric light curves of RX J0859 are shown in Figure \ref{fig:optlc_rxj08}.   Light curves of each epoch of observations exhibit a clear eclipse profile where the brightness of the system dropped by an average of $\sim$ 1.1 mag. To know the orbital period and the eclipse morphology of RX J0859, we determined mid-eclipse times (see Table \ref{tab:eclipsetime_07_08}) by fitting parabolas to the bottom part of eclipses for all epochs of observations except one - the epoch of 03 May 2014 (where the whole eclipse profile was not observed). A linear least square fit to the seven eclipse timings provides the following ephemeris for RX J0859:
\begin{equation}                         
T_0 = HJD 2456688.4296 (19) \pm 0.0999294 (6) E,
\label{eq:eq1}
\end{equation}

\noindent
where T$_0$ is defined as the time of mid-eclipse and the errors are given in brackets. 

The orbital period of RX J0859 was thus derived to be 2.39 hrs. A careful visual inspection showed that the primary eclipse going to complete one cycle during  27 March 2015 and 09 April 2015 provided an estimate of the orbital period of $\sim$ 2.4 hrs. We have also performed the period analysis of RX J0859 by applying Fourier Transform (FT) to the photometric data and using the Lomb-Scargle periodogram algorithm \citep[see][]{Lomb76, Scargle82, Horne86}. The Lomb-Scargle power spectrum of RX J0859 is shown in the top panel of Figure \ref{fig:optps_rxj08}. Various peaks are observed in the power spectrum as the fundamental and harmonics of the orbital period. These peaks are marked as $\Omega$, $2\Omega$, and $3\Omega$. The errors on these periods are derived by using the method given in \citep{Horne86}. The significance of this detected peaks is determined by calculating the false alarm probability \citep[see][]{Horne86}. The horizontal line shows a 99\% significance level. The Scargle power spectrum of RX J0859 is noisier due to data gaps. Therefore, true variations in the source are further modulated by the irregular sampling defined by the window function of the data. We have computed the window function with the same time sampling but contains only a constant magnitude and is shown in the middle panel of Figure \ref{fig:optps_rxj08}. The frequencies identified in the  Lomb-Scargle power spectrum did not fall under the window function, which assures that the periods derived from the Lomb-Scargle power spectrum are real. In order to further confirm the periodicity of this system, the {\sc clean} algorithm \citep{Roberts87} is applied to the data. Lower panel of Figure \ref{fig:optps_rxj08} shows the CLEANed power spectrum of all photometric data of RX J0859. The CLEANed power spectrum was obtained with a loop gain of 0.1 and 1000 iteration. Similar to the Scargle periodogram, the peaks corresponding to the frequencies $\Omega$, $2\Omega$, and $3\Omega$ are also found in the CLEANed power spectrum. The error in the period obtained from {\sc clean} method is determined as P$^2$/2t$_{max}$, where P is the period and t$_{max}$ is the duration of the observations \citep[see][]{Roberts87}. The periods corresponding to these frequencies are given in Table \ref{tab:ps}. These power spectral analysis further confirms that the derived orbital period of RX J0859 is 2.393$\pm$0.003 hrs.

The photometric light curves of all these seven observations were also folded using the ephemeris given in equation \ref{eq:eq1} and are shown in Figure \ref{fig:optfoldlc_rxj08}. Most of the light curves show a V-shape eclipse profile, whereas the light curve of 27 March 2015 exhibits a relatively flat bottom of the eclipse. 

The mass (M$_2$) and radius (R$_2$) of the secondary are estimated to be 0.19$\pm$0.07$M_{\odot}$ and 0.24$\pm$0.03$R_{\odot}$, respectively  with the help of  the mean empirical mass-period relation of \citet{Smith98}. We have estimated the mean density ($\overline{\rho}$) of secondary as 18.6$\pm$0.1 g cm$^{-3}$ \cite[see][for formulae]{Warner95}. This mean density of secondary corresponds to a lower main sequence star of spectral class M4 and an effective temperature of 3300 K \citep[see][]{Beuermann98, Knigge06}.
 Using the mass range of WD of 0.2-1.5 $M_{\odot}$ in CVs \citep{Zorotovic11}, the binary separation (a) is estimated to be in the range of ($4.6 - 7.5$)$\times10^{10}$ cm. 

\begin{figure}
\centering
\hspace{-0.5 cm}
\includegraphics[width=90mm]{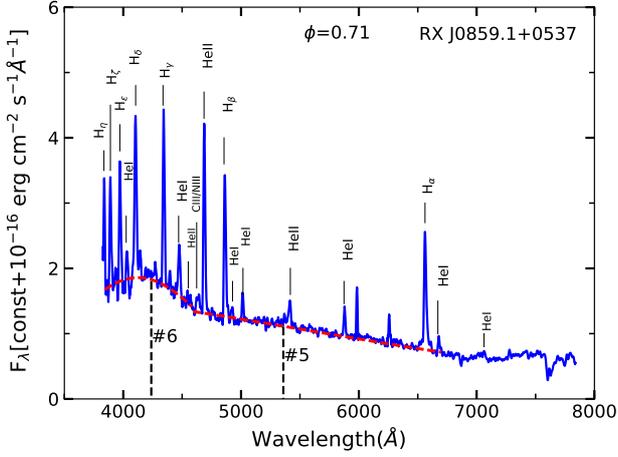}
\caption{Optical spectrum of RX J0859.1+0537. The orbital  phase of the observation is mentioned at the top of the spectrum. The overlaid solid dash red lines are the best fitted Gaussian to the cyclotron hump after excluding the emission lines. Vertical dashed lines represent the cyclotron humps and corresponding cyclotron harmonic numbers are also mentioned.}	
\label{fig:optspec_rxj08}
\end{figure}

\subsubsection{Optical Spectroscopy}
\label{sec:optspec08}
A very rich optical spectrum with a wealth of spectral features was obtained at out-of-eclipse phase $\phi$=0.71 for RX J0859 and is shown in Figure \ref{fig:optspec_rxj08}.  Identification, fluxes, equivalent width (EW), and FWHM of the principal emission lines of RX J0859 obtained through a single-Gaussian fitting are given in Table \ref{tab:opt_spec_parm}. The optical spectrum is dominated by the single-peaked Balmer emission lines (from H$\alpha$ to H$\eta$). The high-excitation line He II 4686 \AA ~is quite prominent and its strength is 6/5 of the H$\beta$-flux. Several emission lines of He I, He II, and the Bowen fluorescence lines are also present in the optical spectrum of RX J0859.  The Bowen fluorescence line is observed with the NIII line along with an additional contribution from CIII. In addition to He II 4686 \AA ~line, we have also found a weak He II 4542 \AA ~emission line from the Pickering series. \cite{Schachter91} reported that the Pickering series is also observed in few AM Her systems. The flux ratio of He II 4542 \AA ~to He II 4686 \AA ~is found to be 0.1. 

Two broad features are clearly observed in the optical spectrum of RX J0859 and are identified as due to cyclotron emission that occurs at discrete harmonics of cyclotron frequency. Unlike the first hump, the second hump appears to be shallower in this system.  These humps are equally spaced in frequency for a given electron at low energies where the relativistic effects are not important. The central wavelength of each cyclotron hump was estimated by fitting the Gaussian curve. The error  corresponds to the central wavelength of each hump is  the standard deviation obtained from the Gaussian fitting. Any measurement of the position of the cyclotron hump is problematic by the intense emission lines in the spectrum. Thus, the emission lines were removed while fitting the Gaussian to the hump. The dashed line overlaid in the spectrum in Figure \ref{fig:optspec_rxj08} shows the best fit Gaussian curve to the cyclotron humps. We have identified prominent humps at $4238\pm191$ \AA ~and $5359\pm407$ \AA. If we consider these two humps as successive harmonics of fundamental cyclotron frequency ($\nu_{cyc}$), their separation provides the measured humps at  $4238$ \AA ~and  $5359$ \AA  ~as the 6$^{th}$ and   5$^{th}$ harmonics of fundamental cyclotron frequency, respectively. 
 For a given magnetic field, the fundamental cyclotron frequency is described as  $\nu_{cyc}$ =  2.8$\times$10$^{12}$ B$_6$ Hz or $\omega_{B}$ = 2 $\pi$ $\nu_{cyc}$ = 1.76 $\times$10$^{13}$ $B_{6}$ \citep[see][]{Ingham76, Warner95}. However, the exact value of the magnetic field will, in fact be, depend on the temperature of the plasma and the viewing angle relative to the field lines \citep[for more detail see][]{Wickramasinghe82, Barrett85}. Here, $B_6$ is magnetic field in terms of 10$^6$ G. Using the value of $\omega_{B}$, the equation (3) of \cite{Cropper89} can be written  as  

\begin{equation}
B_{6} =  \frac {4.28 \times 10^{6} sin^{2}\theta} {\mu \lambda \left[{-1+\sqrt{1+{(8nsin^{2}\theta}/\mu)} }\right ]}  
\label{eq:eq2}
\end{equation}
where $\lambda$ is in the unit of \AA, $\mu$ = $m_{e}$c$^2/kT$ = $511.1/T$, $T$ is the plasma temperature in the unit of keV, $n$ is the cyclotron harmonic number, and $\theta$ is the viewing angle to the magnetic field line. In order to determine the magnetic field strength, we have tried to constrain the temperature T and the viewing angle $\theta$ for which the consistent magnetic field for both observed harmonics can be determined. For T=10 keV and $\theta$=90$^{\circ}$, we have estimated the consistent magnetic field of 50$\pm$2 MG and B=47$\pm$4 MG for the 6$^{th}$ and 5$^{th}$ harmonics, respectively. These values of the estimated magnetic field were well within the $1\sigma$ level. The uncertainty on B was derived from uncertainty in the central wavelength of cyclotron harmonics. Thus, the average value of the magnetic field strength of the WD in the system RX J0859 is estimated to be $49\pm2$ MG.

\begin{table}
\caption{Identification, flux, EW, and FWHM for emission features of RX J0859.1+0537, RX J0749.1-0549, and RX J0649.8-0737.\label{tab:opt_spec_parm}}
\tiny
\setlength{\tabcolsep}{0.02in}
\begin{tabular}{cccccccccccccc}
\hline\hline
{\bf Identification} &&\multicolumn{3}{c}{\textbf{RX J0859.1+0537}} && \multicolumn{3}{c} {\textbf{RX J0749.1-0549}}&& \multicolumn{3}{c} {\textbf{RX J0649.8-0737}}\\
\cline{3-5} \cline{7-9} \cline{11-13}
                     &&Flux&$-$EW&FWHM&&Flux&$-$EW&FWHM&&Flux&$-$EW&FWHM\\
\hline\hline
H$_\eta$ (3835~\AA)       && 1.36 & 7  &1839 &&1.26&7    &1026&&3.25&7 &1281\\
H$_\zeta$ (3889~\AA)      && 2.83 & 17 &1990 &&1.50&10   &1152&&4.79&11&1255\\
H$_\epsilon$ (3970~\AA)   && 3.15 & 18 &1357 &&2.65&20   &1194&&6.62&15&1369\\
HeI (4026~\AA)            && 1.22 & 7  &1482 &&0.47&3    &1640&&1.11&3 &1104\\
H$_\delta$ (4102~\AA)     && 4.87 & 25 &1678 &&9.39&53   &949 &&8.15&16&1392\\
H$_\gamma$ (4340~\AA)     && 4.75 & 27 &1417 &&3.51&19   &1322&&9.86&20&1319\\
HeI (4471~\AA)	          && 2.01 & 13 &1802 &&1.28&8    &2122&&3.09&7 &1529\\ 
HeII (4542~\AA)	          && 0.30 & 2  &925  &&0.31&2    &1752&&1.42&5 &2591\\ 
CIII/NIII (4640/50~\AA)   && 0.49 & 4  &1872 &&0.41&1    &1229&&1.05&4 &2500\\ 
HeII (4686~\AA) 	  && 4.96 & 36 &1296 &&1.37&9    &1853&&5.93&14&1122\\ 
H$_\beta$ (4861~\AA)      && 4.21 & 34 &1348 &&3.01&22   &1318&&9.76&24&1287\\
HeI (4922~\AA)	      	  && 0.42 & 4  &1304 &&0.31&2    &1407&&1.27&3 &1045\\
HeI (5016~\AA)	          && 0.62 & 5  &1060 &&$..$&$..$ &$..$&&1.03&3 &837\\ 
FeII (5169~\AA)	          && $..$ &$..$&$..$ &&0.40&3    &1854&&1.06&3 &1568\\ 
HeII (5412~\AA)	          && 0.60 & 5  &1179 &&$..$&$..$ &$..$&&1.80&4 &1718\\
HeI (5875~\AA)	          && 0.77 & 8  &1044 &&1.19&9    &1619&&2.28&6 &920\\
H$_\alpha$ (6563~\AA)     && 3.57 & 42 &1093 &&2.32&19   &886 &&9.20&27&971\\
HeI (6678~\AA)	          && 0.58 & 8  &1168 &&1.01&5    &1149&&1.42&5 &988\\
HeI (7065~\AA)	          && 0.64 & 10 &3151 &&0.40&5    &1481&&1.26&5 &1676\\
\hline
\end{tabular}
\vskip.4cm 
{{\bf \it Note.-}} Flux, EW, and FWHM are in the unit of 10$^{-15}$ erg cm$^{-2}$ s$^{-1}$, \AA, and km s$^{-1}$, respectively.
\end{table}

\subsubsection{X-ray Timing Analysis}
\label{sec:x-raylc08}
\begin{figure}
\centering
\includegraphics[width=80mm]{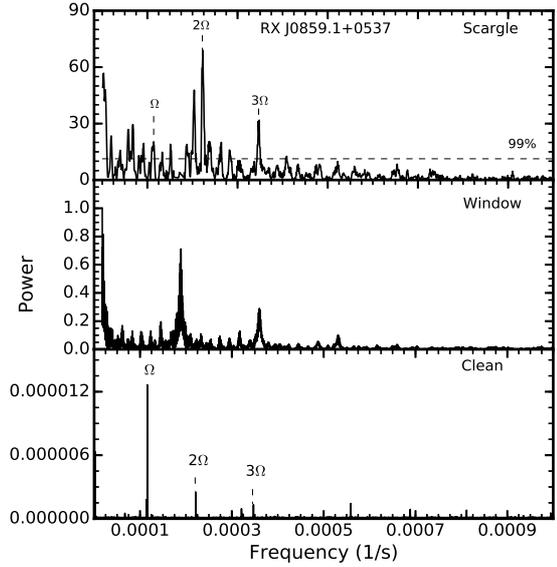}	
\caption{Lomb-Scargle X-ray power spectra ({\it top panel}),  window function ({\it middle panel}), and CLEANed X-ray power spectra ({\it bottom panel}) of RX J0859.1+0537. The horizontal dash line represent 99\% significance level.} 
\label{fig:xrayps_rxj08}
\end{figure}

The background subtracted {\textit ROSAT-HRI} light curve was extracted for one long exposure of RX J0859. Data gaps were seen in the light curve of RX J0859 due to the occultation of the source by the Earth. Periodicity in the X-ray data of RX J0859 was searched by using similar methods as discussed above. The X-ray power spectra of RX J0859 using Lomb-Scargle and {\sc clean} algorithms along with the window function are shown in Figure \ref{fig:xrayps_rxj08}. The highest peaks in both Lomb-Scargle and {\sc clean} power spectra correspond to the orbital period of 2.4 hrs and its harmonics at 1.2 hrs and 0.8 hrs. These values of the period are found to be consistent with those derived from the optical light curves. Also, no window pattern is seen near these frequencies (see also middle panel Figure \ref{fig:xrayps_rxj08}). The X-ray light curve of RX J0859 was also folded with the ephemeris as given in equation \ref{eq:eq1}. The orbital modulation is clearly observed and confirms the orbital period of RX J0859. A double-humped structure is seen in the phased X-ray light curve. The prominent hump is near the orbital phase $\sim$0.5, while the secondary hump is observed at phase $\sim$ 0.85 with a smaller amplitude.

\begin{figure}
\centering
\includegraphics[width=8cm,height=7.8cm, angle=-90]{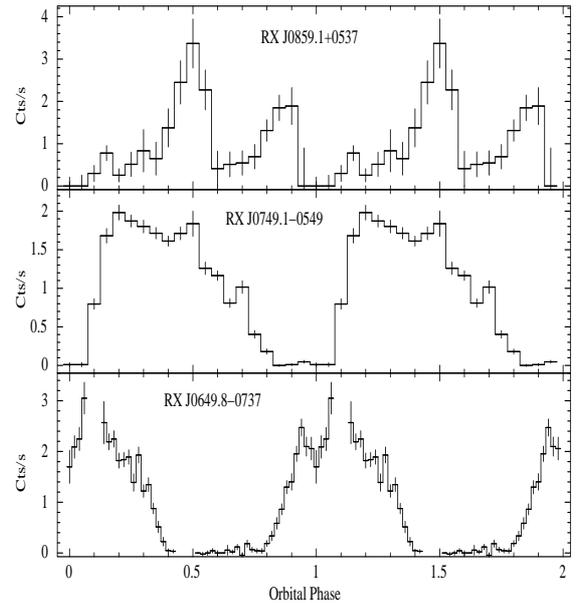}
\caption{Orbital phase folded X-ray light curves of RX J0859.1+0537, RX J0749.1-0549, and RX J0649.8-0737 in energy range 0.1-2.2 keV.}
\label{fig:rosatflc}
\end{figure}

\begin{figure*}
\centering
\subfigure[]{\includegraphics[width=58mm]{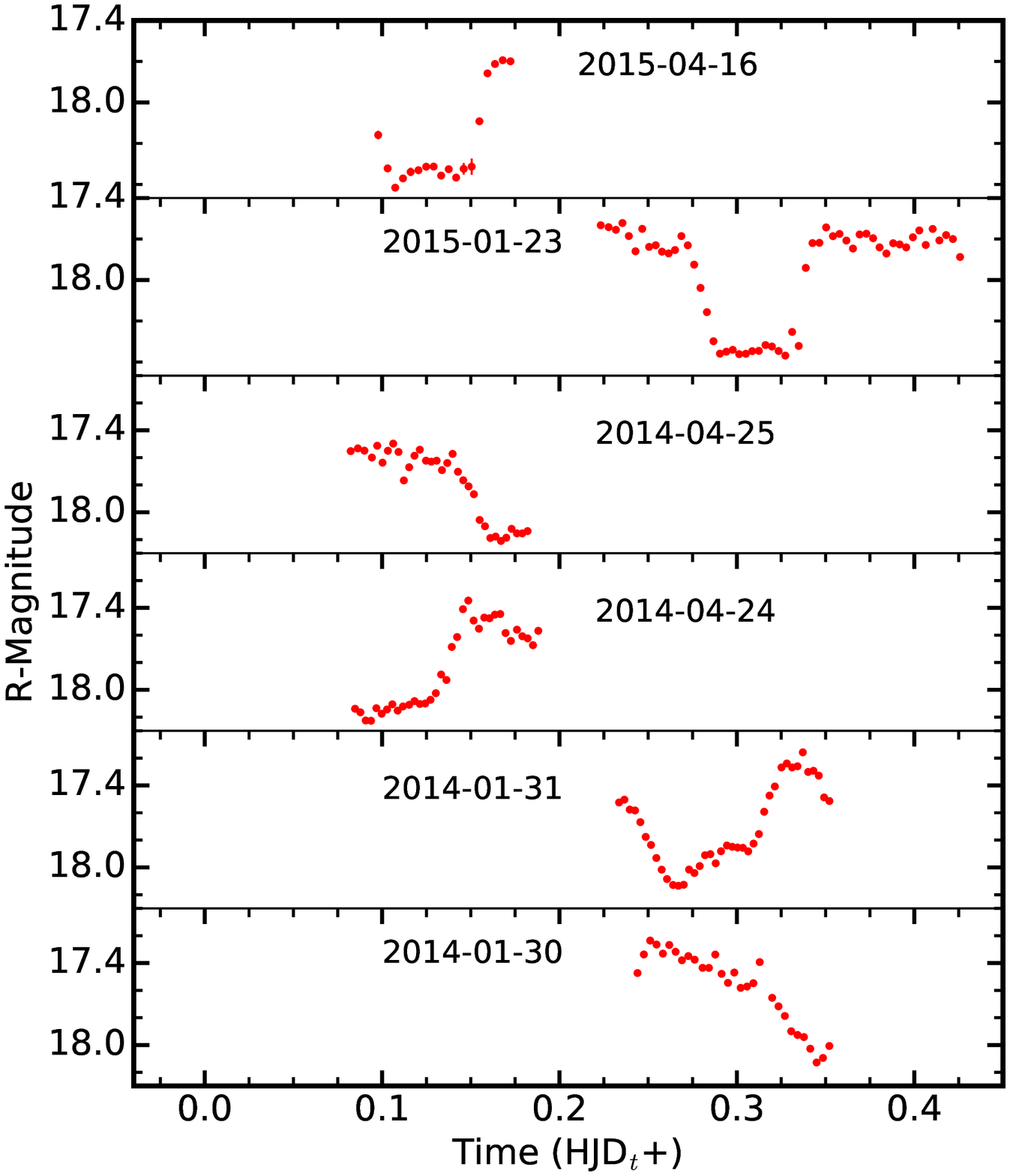}\label{fig:optlc_rxj07}}
\subfigure[]{\includegraphics[height=60mm, width=58mm]{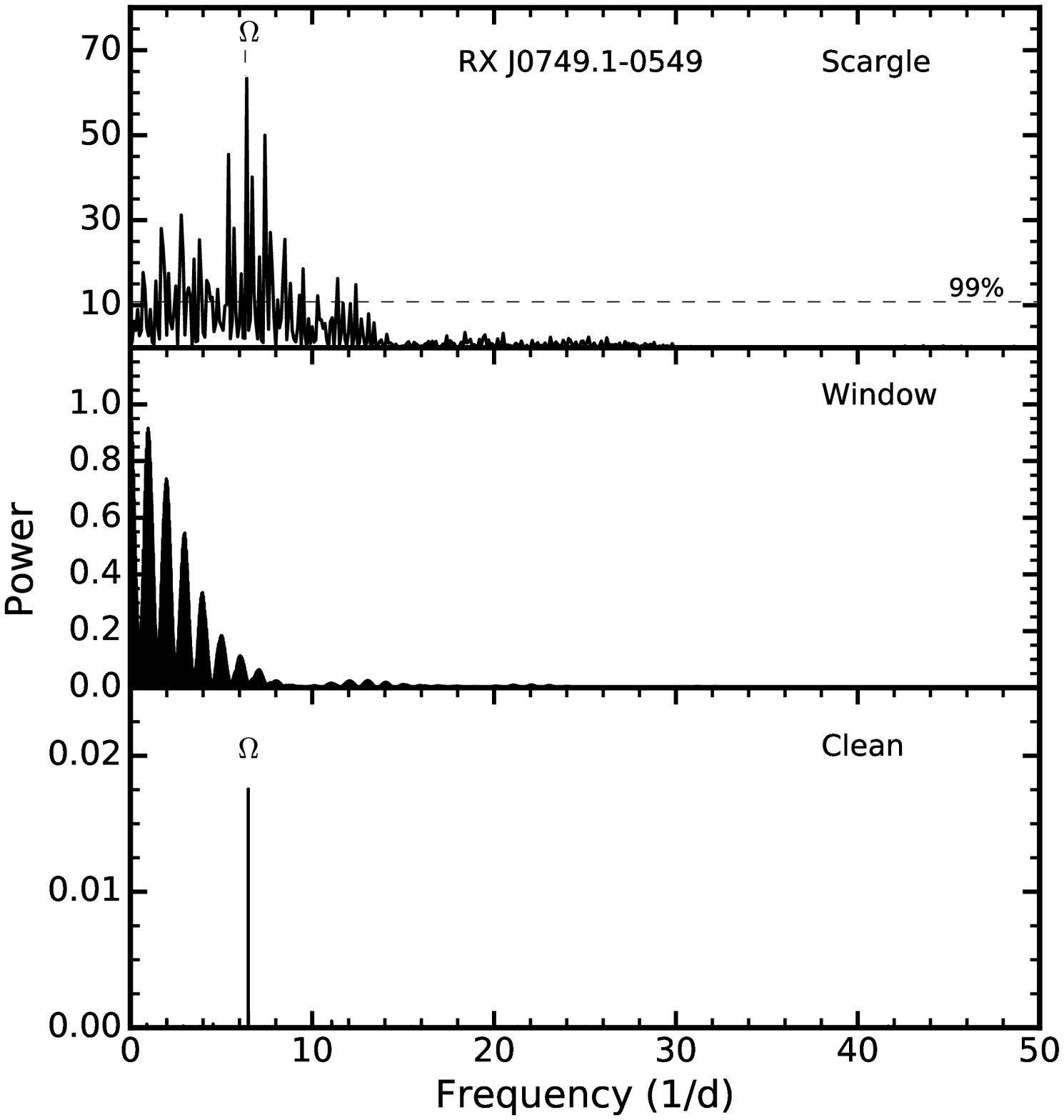}\label{fig:optps_rxj07}}
\subfigure[]{\includegraphics[height=60mm,width=58mm]{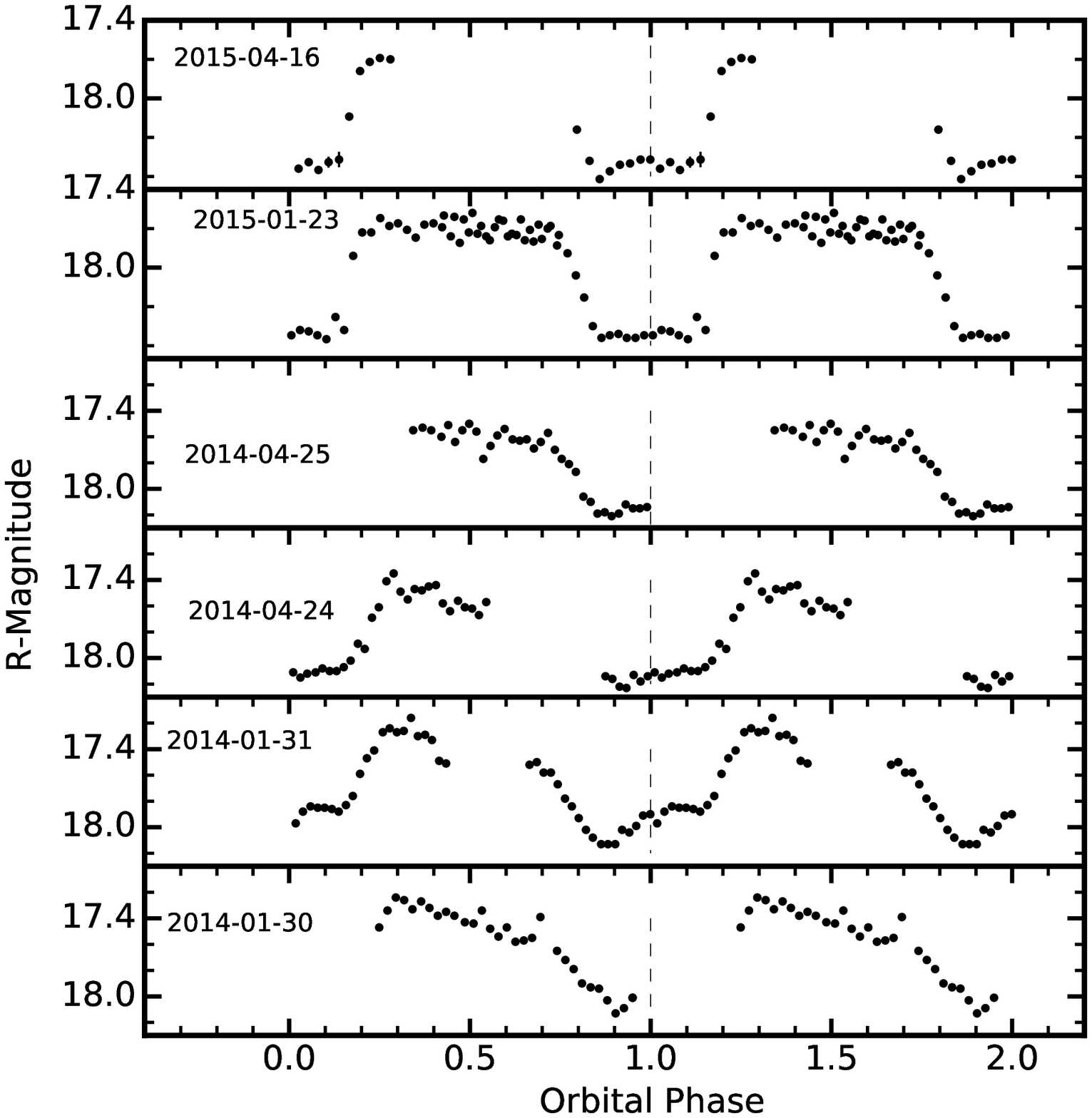}\label{fig:optfoldlc_rxj07}}
\caption{(a) R-band light curve of RX J0749.1-0549, (b) Lomb-Scargle power spectra (top panel), window function (middle panel), and CLEANed power spectra (bottom panel) of RX J0749.1-0549, and (c) folded R-band light curve of RX J0749.1-0549. The date of observations are mentioned near each light curve. The HJD$_{t}$ of each observation  is given in Table \ref{tab:obslog}.} 
\end{figure*}


\subsection{RX J0749.1-0549}
\label{sec:Analysis_RXJ0749}
\subsubsection{Optical Photometry}
\label{sec:phot07}
 RX J0749 was observed for a total of $\sim$ 17 hrs in six nights. Figure \ref{fig:optlc_rxj07} shows the R-band photometric light curves of RX J0749.  Ellipsoidal variations are clearly visible in each epoch of observations. The profiles of eclipses are complicated and highly variable.  We used the mid-eclipse times for all observations, except for 25 April 2014 and 30 January 2014, for the best determination of ephemeris. The entire eclipse profile was not covered in the two observations excluded here.  We also take half the exposure time as the uncertainty in the measurement of mid-eclipse for all four observations. This approach is similar to that mentioned in section \ref{sec:phot08}, and the ephemeris thus determined for the time of mid-eclipse is given as   
\begin{equation}                         
T_0 = HJD  2456688.28521 (3) \pm 0.154225 (2) E .
\label{eq:eq3}
\end{equation}

We thus find that the orbital period of RX J0749 is 3.7014 hrs. A full eclipse was seen during the epoch 23 January 2015 due to the sufficient data length as compared to other epochs of observations. A clear flat bottom was also observed in this observation, where the time duration of total eclipse was $\sim$ 61 min. Power spectra  of RX J0749 are shown in the Figure \ref{fig:optps_rxj07}. The period corresponding to the peak frequency is given in Table \ref{tab:ps}. The {\sc clean} algorithm leads to an orbital period of 3.672$\pm$0.001 hrs, however. As shown in middle panel of Figure   \ref{fig:optps_rxj07}, no window pattern is seen near the identified orbital frequency. The light curves of all six observations were also folded using the ephemeris given in equation \ref{eq:eq3} and are shown in Figure \ref{fig:optfoldlc_rxj07}. Each light curve of RX J0749 shows a noticeable change in the shape of the eclipse profile. The brightness of RX J0749 was found to be decreased from the year 2014 to the year 2015 by $\sim$ 0.4 mag. The eclipse profile was also observed to be more symmetric and flat-bottomed in the year 2015.

The mass and radius of the secondary have been derived to be 0.35$\pm$0.08 $M_{\odot}$ and 0.39$\pm$0.03 $R_{\odot}$, respectively, using the same procedure as above. We have also estimated the separation between the WD and secondary in the range of $(6.88-10.3)$$\times$10$^{10}$ cm assuming M$_1$=$(0.2-1.5)$ $M_{\odot}$. Using the orbital period of 3.672$\pm$0.001 hrs, the mean density of secondary is calculated to be $\overline{\rho}$ =7.940$\pm$0.004 g cm$^{-3}$ which indicates that the secondary of RX J0749 has a spectral type of M3.7 and an effective temperature of 3400 K.

\subsubsection{Optical spectroscopy}
\label{sec:optspec07}

\begin{figure}
\centering
\hspace{-0.5 cm}
\includegraphics[width=90mm]{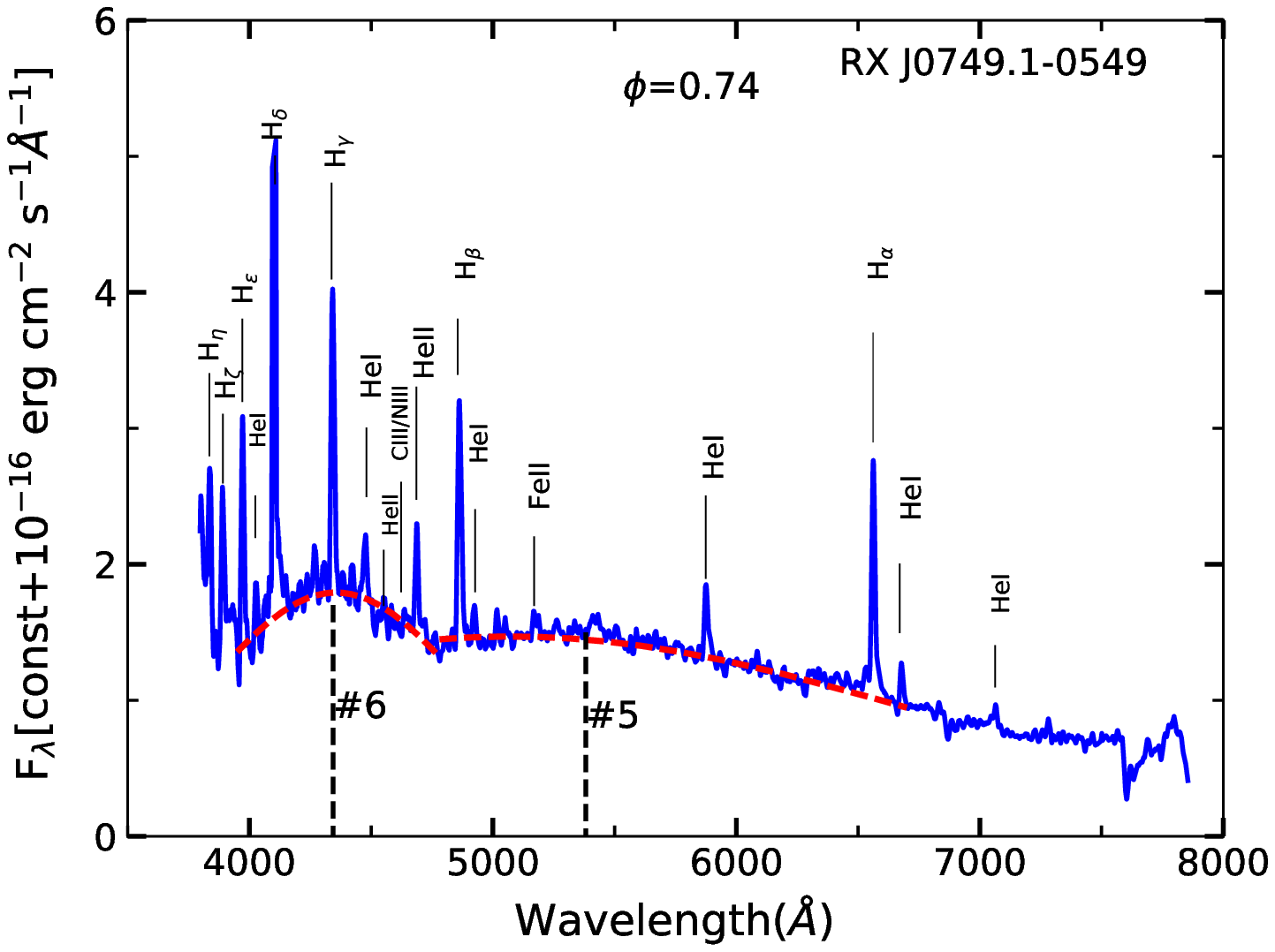}
\caption{Optical spectrum of RX J0749.1-0549. The orbital  phase of the observation is mentioned at the top of the spectrum. The overlaid solid dash red lines are the best fitted Gaussian to the cyclotron hump after excluding the emission lines. Vertical dashed lines represent the cyclotron humps and corresponding cyclotron harmonic numbers are also mentioned.}	
\label{fig:optspec_rxj07}
\end{figure}

Low-resolution spectrum of RX J0749 was obtained at an orbital phase of 0.74 and is shown in Figure \ref{fig:optspec_rxj07}. The spectrum is dominated by the Balmer emission lines, He II 4686 \AA ~emission features along with the lines of He I, He II, and the Bowen CIII/NIII. In contrast to RX J0859, the strength of the He II 4686 \AA ~emission line is relatively weak and corresponds to 7/5 of the H$\beta$-flux. The emission line of Fe II 5169 \AA ~is also visible in the spectrum of RX J0749. The flux ratio of He II 4542 \AA ~to He II 4686 \AA ~is found to be $\sim$ 0.2. The parameters derived from the optical spectrum of RX J0749 are given in Table \ref{tab:opt_spec_parm}.

As in the case of RX J0859, two cyclotron humps with variable amplitude are also seen in the optical spectrum of RX J0749. Following the method described above, the central wavelengths of these humps were found to be  $4344\pm171$ \AA ~and $5382\pm279$ \AA ~and are identified as 6$^{th}$ and 5$^{th}$ harmonics of a fundamental cyclotron frequency, respectively.   Considering T of 9 keV and $\theta$ of  40$^{\circ}$, comparable  magnetic field strengths of 44$\pm$2 MG for 6$^{th}$ harmonic and 43$\pm$2 MG for 5$^{th}$ harmonic were found. This  reveals an average field strength of 43.5$\pm$1.4 MG for RX J0749.

\subsubsection{X-ray Timing Analysis}
\label{sec:x-raylc07}
\begin{figure}
\centering
\includegraphics[width=80mm]{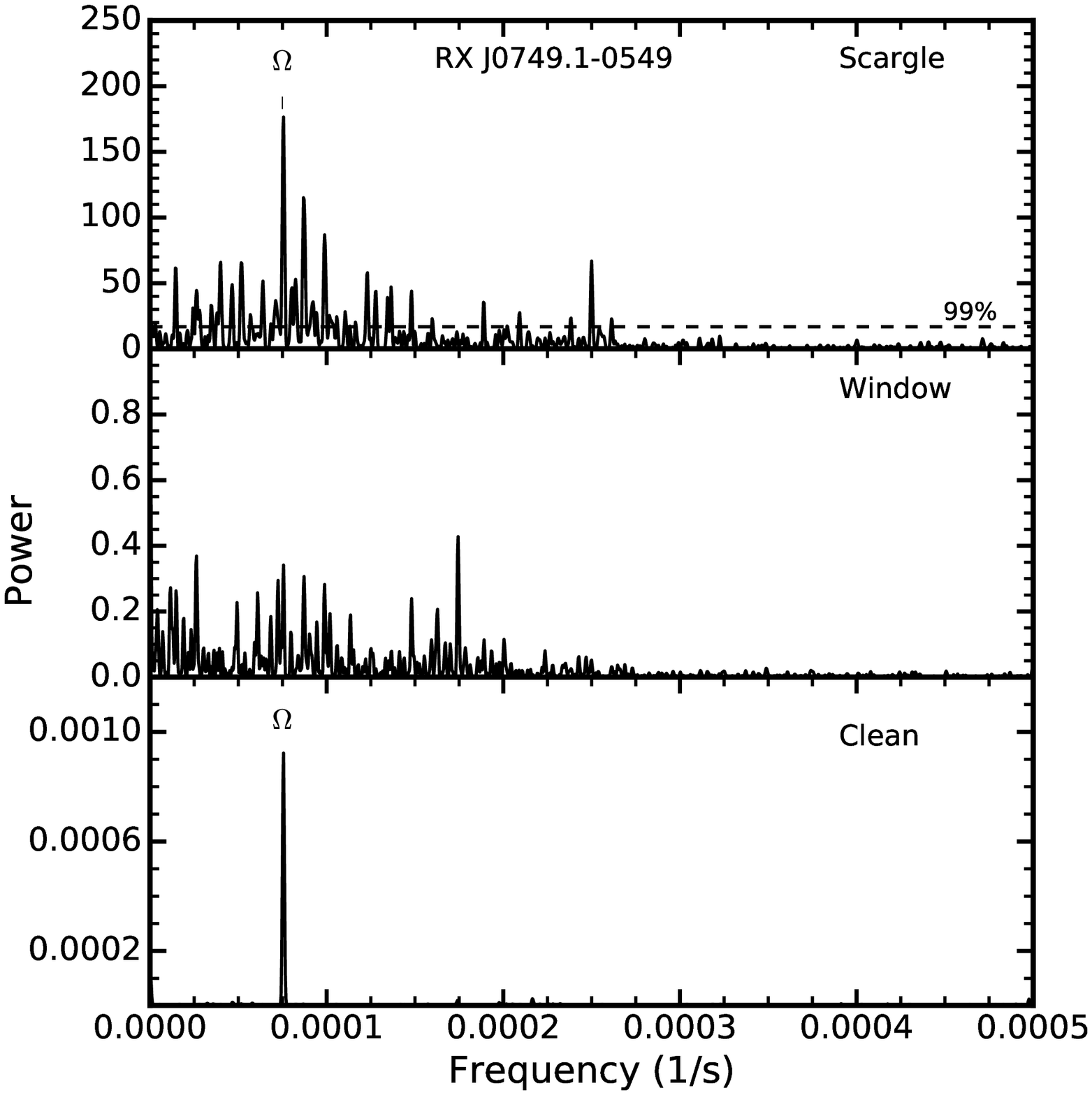}
\caption{Lomb-Scargle X-ray power spectra ({\it top panel}),  window function ({\it middle panel}), and CLEANed X-ray power spectra ({\it bottom panel}) of RX J0749.1-0549. The horizontal dash line represent 99\% significance level.} 
\label{fig:xrayps_rxj07}	
\end{figure}

The power spectra of the X-ray light curve of  RX J0749 as obtained from Lomb-Scargle and {\sc clean} periodograms  are shown in the top and bottom panels of Figure \ref{fig:xrayps_rxj07}. The highest peak in both Lomb-Scargle and CLEANed power spectra corresponds to the orbital period of $\sim$ 3.68 hrs, which is similar to that derived from the R-band light curve (see Table \ref{tab:ps}). As shown in the middle panel of Figure  \ref{fig:xrayps_rxj07}, no windowing is seen near the significant peak in the Lomb-Scargle power spectrum.
 The X-ray light curve was folded with the derived ephemeris as given in equation \ref{eq:eq3}. The folded light curve  is shown in the middle panel of the Figure \ref{fig:rosatflc}. The orbital modulation appears to be maximum around phases $\sim$ 0.25 and interrupted by a broad, eclipse-like feature centered around the orbital phase 1.0, as seen in the R-band light curve.


\subsection{RX J0649.8-0737}
\label{sec:Analysis_RXJ0649}
\subsubsection{Optical Photometry}
\label{sec:phot06}
\begin{figure*}
\centering
\subfigure[]{\includegraphics[height=60mm, width=58mm]{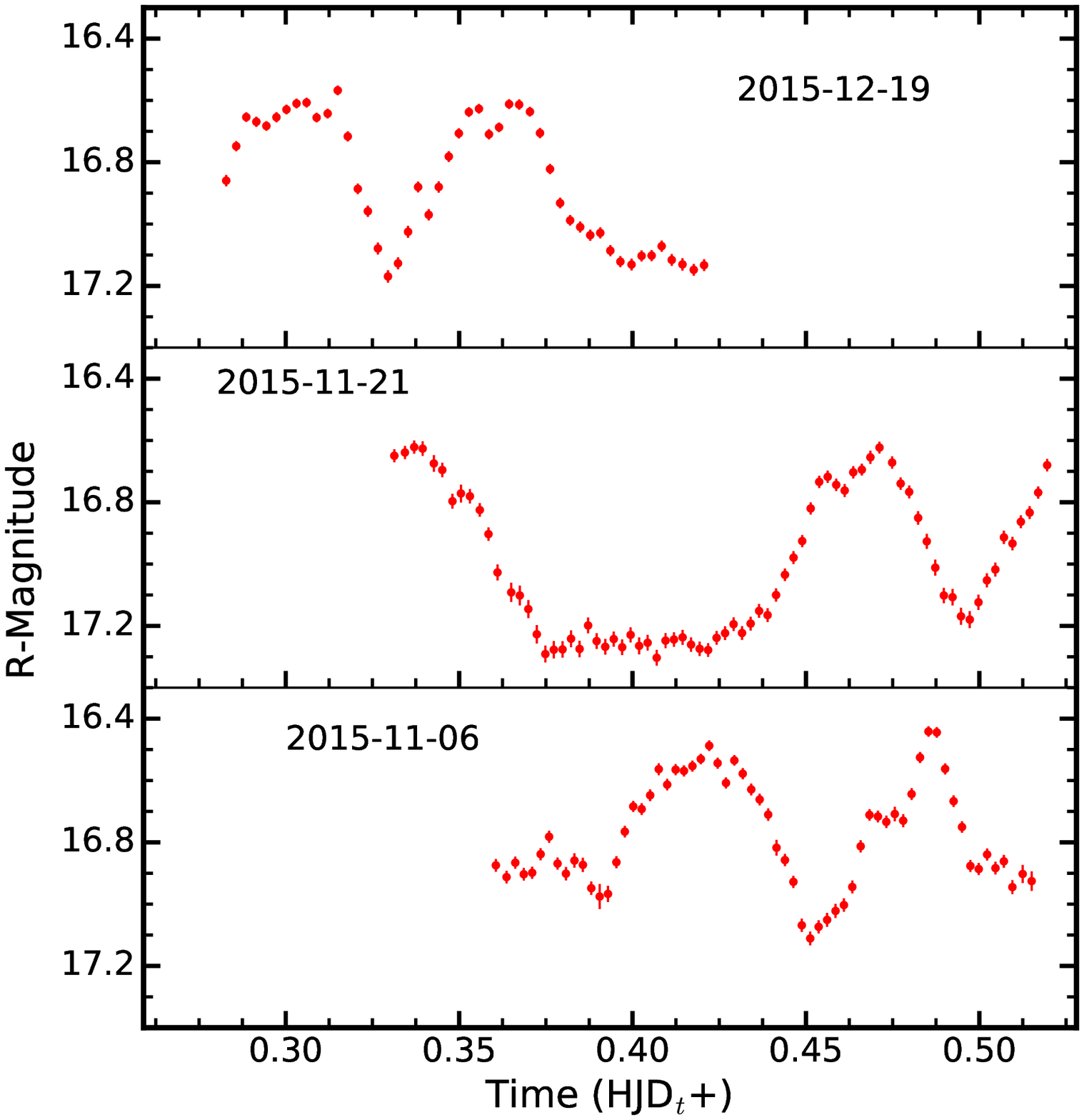}\label{fig:optlc_rxj06}}
\subfigure[]{\includegraphics[height=60mm, width=58mm]{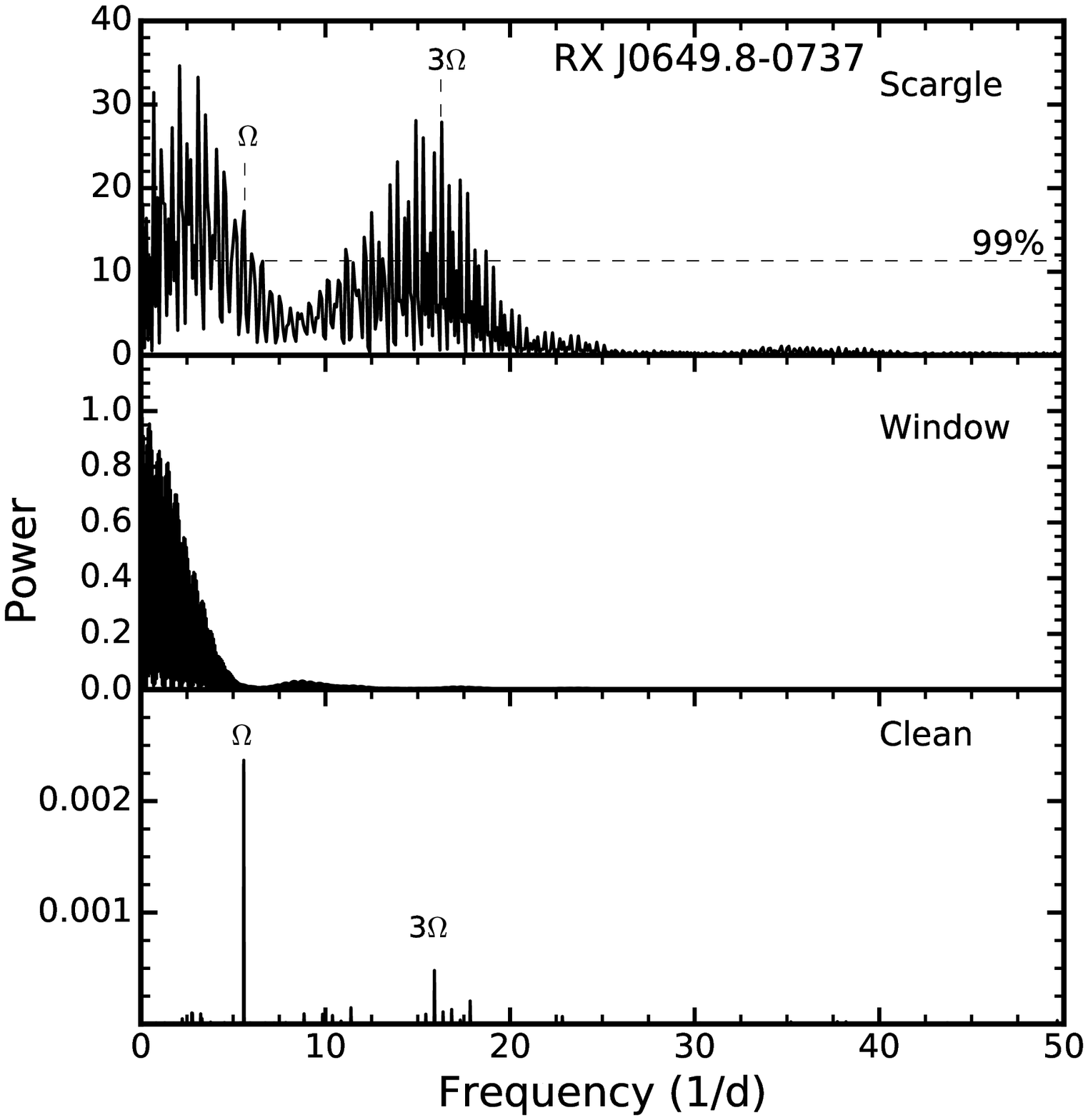}\label{fig:optps_rxj06}}
\subfigure[]{\includegraphics[height=60mm,width=58mm]{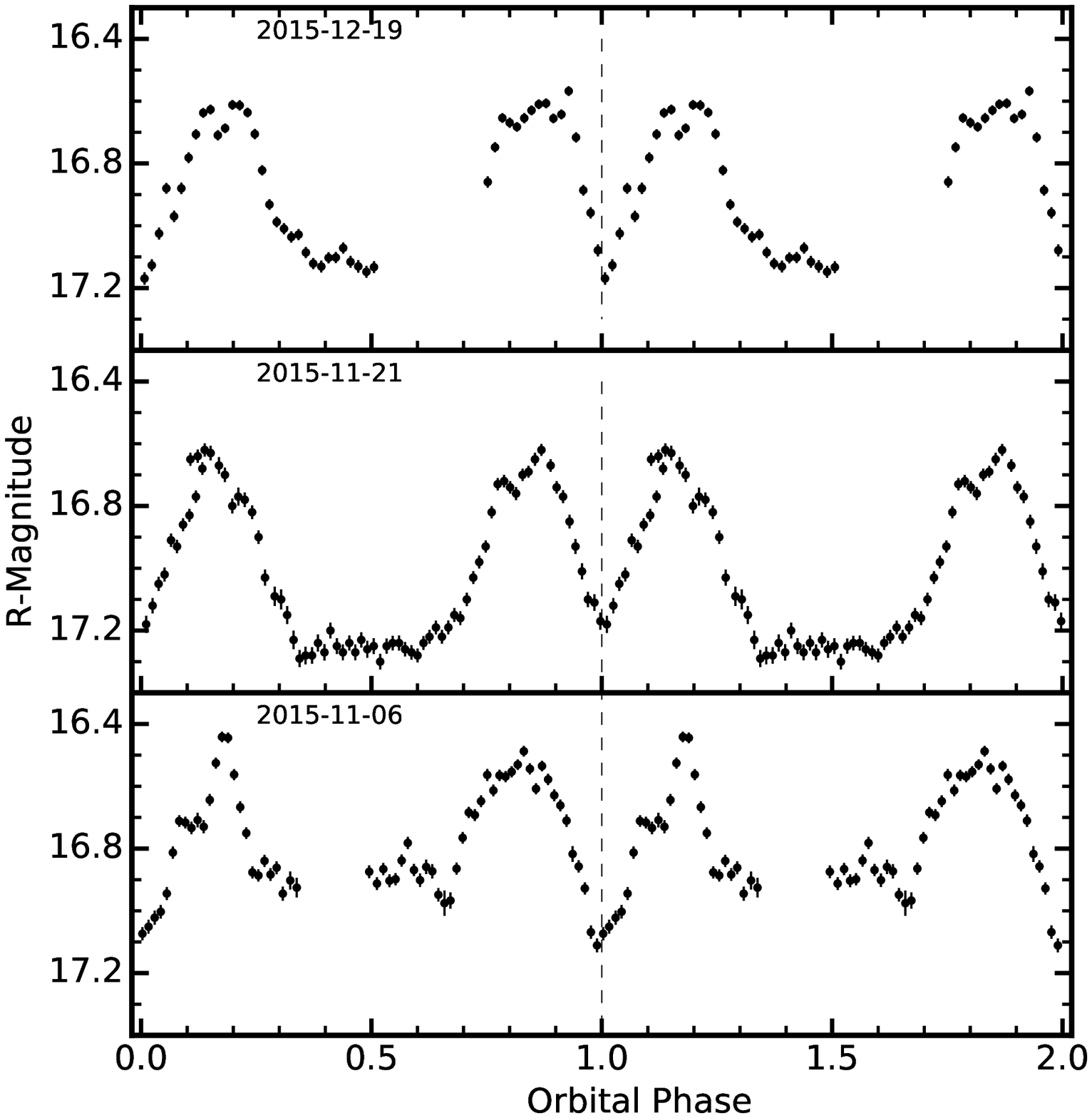}\label{fig:optfoldlc_rxj06}}
\caption{(a) R-band light curve of RX J0649.8-0737, (b) Lomb-Scargle power spectra (top panel), window function (middle panel), and CLEANed power spectra (bottom panel) of RX J0649.8-0737, and (c) folded R-band light curve of RX J0649.8-0737. The date of observations are mentioned near each light curve. The HJD$_{t}$ of each observation  is given in Table \ref{tab:obslog}.} 
\label{fig:optlc}
\end{figure*}

The R-band light curves of  RX J0649  are shown in Figure \ref{fig:optlc_rxj06} for three sets of observations. A total of 11.5 hrs of data were obtained for this object. The ephemeris for the time of mid-eclipse (see Table \ref{tab:eclipsetime_07_08}) deduced from the three observations is
\begin{equation}                         
T_0 = HJD 2457333.451484 (20) \pm 0.183337 (4) E.
\label{eq:eq4}
\end{equation}
 Lomb-Scargle, window function, and CLEANed power spectra of the R-band  light curves of this system are shown in Figure \ref{fig:optps_rxj06}. A period of $4.310\pm0.002$ hrs is found corresponding to the highest peak in the CLEANed power spectrum, which is close to that seen in the Lomb-Scargle power spectrum. This peak is also  found to be free  from any spectral window.  We interpret this period as the orbital period of the system. The significant second harmonic of the orbital period is also detected in both power spectra. The light curves for all epochs of observations were folded using the period of 4.34 hrs and are shown in Figure \ref{fig:optfoldlc_rxj06}. For each observation, the optical variations show a well-defined and deep eclipse-like feature centered around the orbital phase $\sim$ 0.0 with an amplitude of $\sim$ 0.6 mag. Additionally, a prominent feature is present in the form of double-hump peaking at about phase $\sim$ 0.2 and $\sim$ 0.8. Both the humps are found to be equally prominent.

Using a similar approach as mentioned for the other two sources, we derive the mass and radius of the secondary as 0.43$\pm$0.08 $M_{\odot}$ and 0.46$\pm$0.04 $R_{\odot}$, respectively.  The mean density of 5.662$\pm$0.003 g cm$^{-3}$ for the secondary of RX J0649 indicates that it has a spectral type of M3.2 and an effective temperature of 3500 K.
 Considering the  value of $q$ in the range of $2.15-0.29$, the binary separation  is  estimated to be in the range of $(8.06-11.7)$$\times$10$^{10}$ cm.

\subsubsection{Optical spectroscopy}
\label{sec:optspec06}
\begin{figure}
\centering
\hspace{-0.5 cm}
\includegraphics[width=90mm]{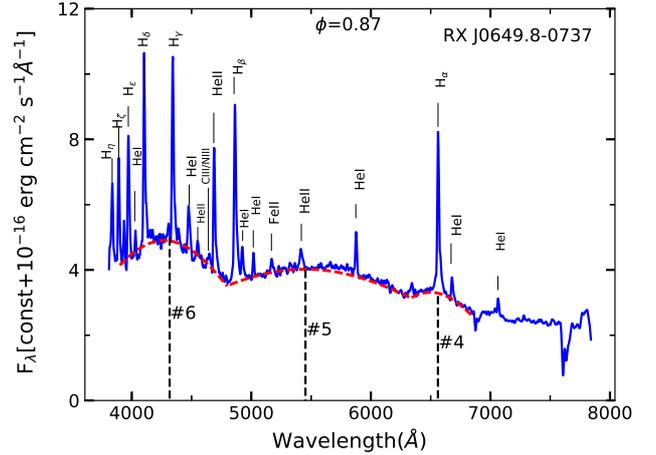}
\caption{Optical spectrum of RX J0649.8-0737. The orbital  phase of the observation is mentioned at the top of the spectrum. The overlaid solid dash red lines are the best fitted Gaussian to the cyclotron hump after excluding the emission lines. Vertical dashed lines represent the cyclotron humps and corresponding cyclotron harmonic numbers are also mentioned.}	
\label{fig:optspec_rxj06}
\end{figure}

The optical spectrum of RX J0649 was obtained at the orbital phase of 0.87 and is shown in Figure \ref{fig:optspec_rxj06}. Similar to the other two spectra, the spectrum of RX J0649 also shows the strong single peaked Balmer, He II, He I, and CIII/NIII emission lines on top of a blue and red continuum. In contrast to RX J0859 and RX J0749, the Bowen blend and Fe II emission lines are found to be stronger in the optical spectrum of RX J0649. The strength of the He II 4686 \AA ~emission line is 3/5 of the H$\beta$-flux. For RX J0649, the flux ratio of He II 4542 \AA ~to He II 4686 \AA ~is similar to as RX J0749. In the Table \ref{tab:opt_spec_parm}, we list the fluxes, EW, and FWHM of these prominent emission lines.

Three broad spectral humps are found to be present in RX J0649. These humps are found to be more prominent as compared to the humps present in the spectrum of RX J0859 and RX J0749. Using a similar approach as mentioned above for two sources, the central wavelength of these humps were estimated to be  $4317\pm219$ \AA, $5453\pm319$ \AA, and $6561\pm121$ \AA ~as 6$^{th}$, 5$^{th}$, and 4$^{th}$ harmonics, respectively. If we consider  T = 11 keV and $\theta$ = 50$^{\circ}$, the magnetic field strengths were found to be  more consistent for each harmonic. We found the magnetic field strength of  45$\pm$2 MG, 43$\pm$2 MG, and 44$\pm$1 MG for 6$^{th}$,  5$^{th}$, and 4$^{th}$ harmonics of cyclotron frequency, respectively. Thus, the average value of the magnetic field strength of the WD is estimated to be 44$\pm$1 MG for RX J0649.

\subsubsection{X-ray Timing Analysis}
\label{sec:x-raylc06}
\begin{figure}
\centering
\includegraphics[width=80mm]{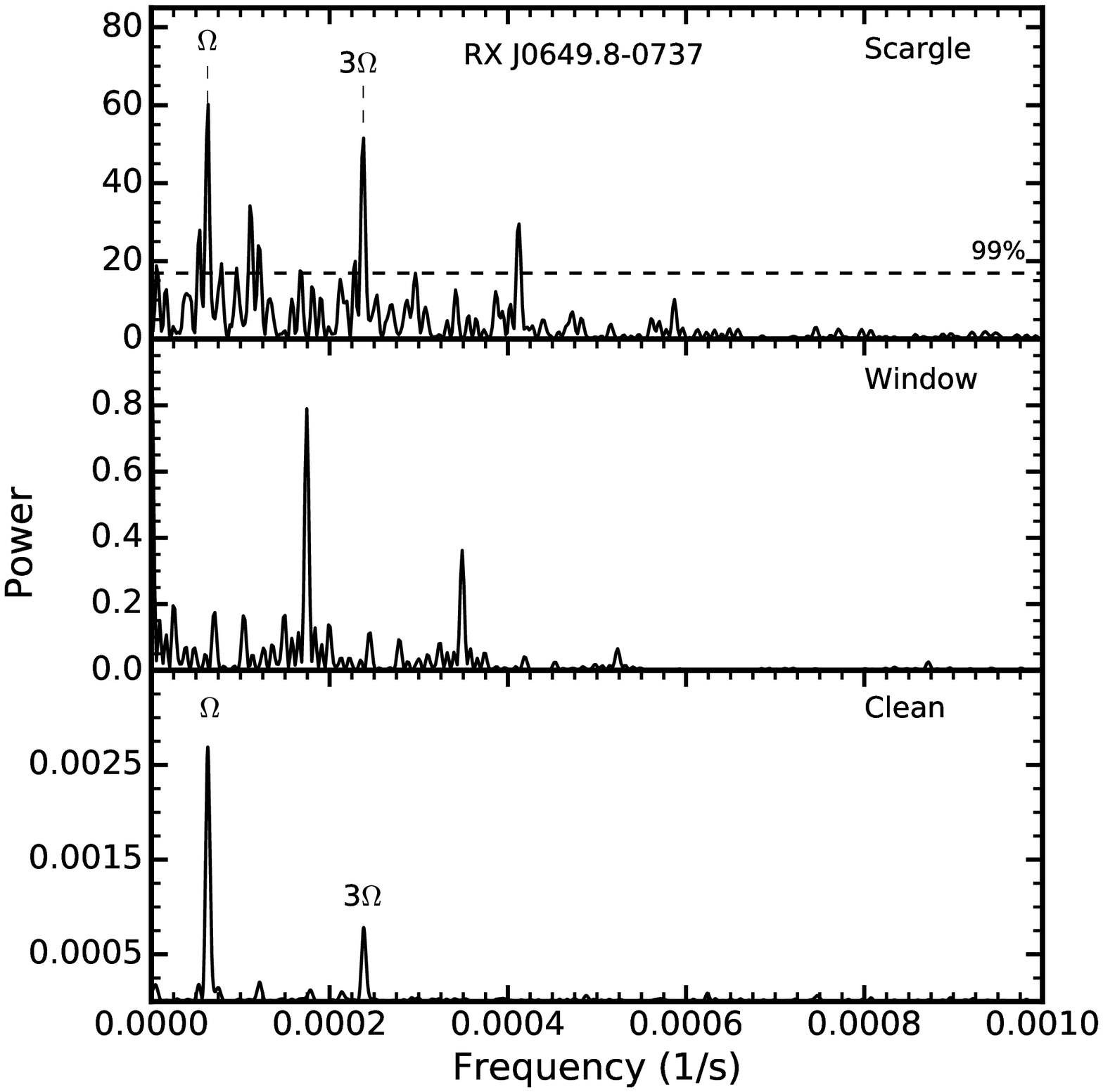}	
\caption{Lomb-Scargle X-ray power spectra ({\it top panel}),  window function ({\it middle panel}), and CLEANed X-ray power spectra ({\it bottom panel}) of RX J0649.8-0737. The horizontal dash line represent 99\% significance level.} 
\label{fig:xrayps_rxj06}
\end{figure}

Figure \ref{fig:xrayps_rxj06} shows the  power spectra of the X-ray light curve of RX J0649. The X-ray power spectra of this system  has also revealed a significant period corresponding to the orbital period of 4.34 hrs. Similar to the R-band power spectra, a second harmonic of the orbital period is also seen in the X-ray power of RX J0649. As seen from the middle panel of Figure \ref{fig:xrayps_rxj06}, this significant peak was free form any spectral window. The modulation found in the X-ray data is displayed by folding the {\textit ROSAT-HRI} light curves using the ephemeris given in equation \ref{eq:eq4}, and is presented in the bottom panel of Figure \ref{fig:rosatflc}. The figure shows a strong periodic modulation at the orbital period. Both optical and X-ray light curves of RX J0649 show similar trend along with its orbital phase.  However, unlike the optical data, the X-ray  eclipse was found to be shallow and narrow.


\section{Discussion}
\begin{figure}
\centering
\includegraphics[width=90mm]{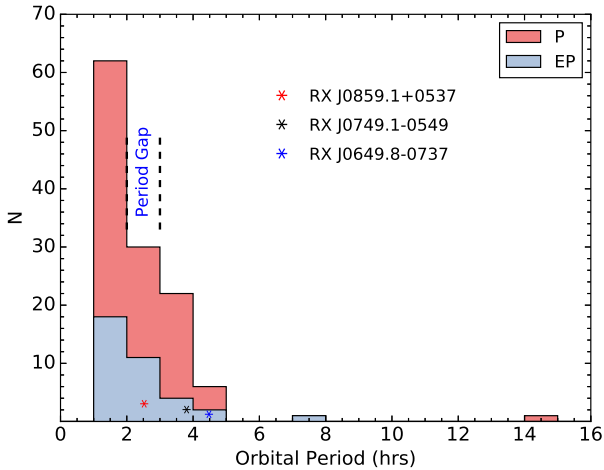}
\caption{The period distribution for confirmed polars (P) and eclipsing polars (EP) taken from RK catalogue. The population of eclipsing polars also includes the newly identified eclipsing polars in this study and are represented by symbol stars.}
\label{fig:perdistpolars}
\end{figure}

We have carried out detailed optical photometry and spectroscopy, and analysed optical as well as the archival X-ray data of three polars RX J0859, RX J0749, and RX J0649. We show that all these three systems belong to a class of eclipsing polars. Our observations and analysis provide the most accurate determination of the orbital periods for these three polars. The periods derived from R-band photometry are consistent with those derived from X-ray data for all these three systems. The derived orbital period of 2.393$\pm$0.003 hrs for RX J0859 places it in the period gap of orbital period distribution of polars. In contrast to RX J0859, the orbital periods of the other two systems RX J0749 (P$_\Omega$ =3.672$\pm$0.001 hrs) and RX J0649 (P$_\Omega$=4.347$\pm$0.001 hrs) lie above the period gap and are consistent with the previously reported periods by \citet{Motch98}. Till now only five polars namely V895 Cen \citep{Howell97}, MN Hya \citep{Buckley98}, V1309 Ori \citep{Staude01}, V1432 Aql \citep{Littlefield15}, and MLS110213:022733+130617 \citep{Silva15} are found to be eclipsing among all known long period polars. Therefore, RX J0749 and RX J0649 can be placed in the list of long-period eclipsing polars above the period gap along with other known five long period eclipsing polars. The period distribution of polars along with eclipsing polars are shown in Figure \ref{fig:perdistpolars}. In this distribution, we have included all three polars identified eclipsing in this study and other three newly identified eclipsing polars ASASSN-16me, CRTS 035010.7+323230, and 2PBC J0658.0-1746 \citep[see][]{Littlefield16,  Mason19, Bernardini19}. Including these three newly identified polars, the current census of CVs consists of 122 polars out of which only $\sim$30 per cent are identified as eclipsing polars. Whereas, $\sim$ 36 per cent of 30 polars in the period gap and $\sim$ 23 per cent of 30 polars above the period gap are found to have eclipsing nature. Therefore, identification of new eclipsing polars in the different period range is an important addition. Besides the presence of the orbital frequencies, their harmonics are also detected in the power spectrum of RX J0859 and RX J0649. The observed amplitude variations, pulse shaped light curves along with the orbital sideband frequencies could be due to structural changes in the accretion stream around the white dwarf. 

\begin{figure}
\centering
\includegraphics[width=60mm,angle=270]{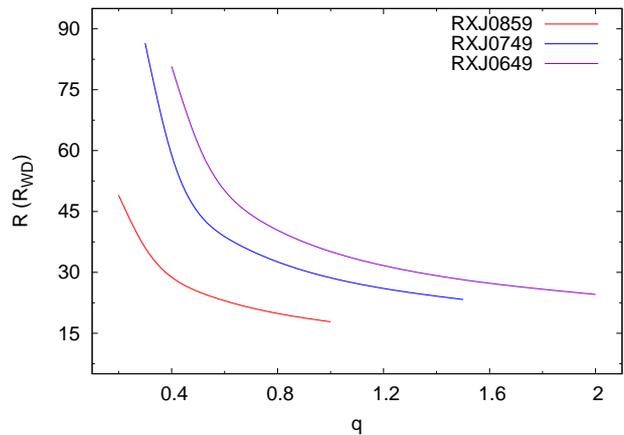}
\caption{The plot of eclipsed size as a function of mass ratio for the polars RX J0859.1+0537, RX J0749.1-0549, and RX J0649.8-0737.}
\label{fig:rq}
\end{figure}

Photometric variations in RX J0859 are found  to be similar to that found in other eclipsing polars like V516 Pup and V834 Cen \citep{Schwarz99, Schwope90}. Light curves of these polars are characterised by V-shaped minimum followed by double humped bright phase. The eclipse observed at phase 0.0 could be associated with the minimum cyclotron flux  when the observer is looking down the accretion column. The radius of eclipsed region ($R$) can be determined by using the following equation \citep{Bailey90}
\begin{equation}
 R =  \pi ~a ~\sqrt{(1-\alpha^2)} ~\Delta \phi_{ie} ,
\label{eq:eq5}
\end{equation}

\noindent 
where $\Delta$$\phi_{ie}$ is ingress/egress duration,  $\alpha$=$cos~i$/$cos~i_{lim}$, $i$ is angle of inclination, and $i_{lim}$ is a limiting angle of inclination for which eclipse half width at half depth reaches zero. The average value of $\Delta$$\phi_{ie}$ for RX J0859 is derived to be $0.21\pm0.05$, where the error on $\Delta \phi_{ie}$ is standard deviation of different measurements. Assuming the angle of inclination, $i \sim 80^\circ$ and using the Table 1 of \cite{Bailey90}, the $R$ as a function of $q$ for all three systems are  plotted in Figure \ref{fig:rq}. Here, $R$ is derived in terms of the radius of WD ($R_{WD}$), where $R_{WD}$ is estimated using the mass-radius relation given by \citet{Nauenberg72}. The limiting values of $q$ are estimated by using the minimum and maximum values of WD in CVs of 0.2 M$_\odot$ and 1.5 M$_\odot$, respectively  \citep[see][]{Zorotovic11}. For the average value of $q$ of 0.24 \citep[assuming the average value of the mass of WD in CVs as 0.8 M$_\odot$;][]{Zorotovic11}, the value of $R$ is estimated to be $\sim$ 40 $R_{WD}$. Even for the maximum value of the q, the value of R is estimated to be more than 18\,R$_{WD}$ for RX J0859. The large value of $R$ indicates that the eclipse observed in the light curves is not solely due to the occultation of WD by secondary. In general, the dimension of the optical and X-ray emission sites are thought to be $\textless$ 0.1 R$_{WD}$ \citep{Bailey91}. Therefore, the observed eclipse in RX J0859 might be due to the occultation of the accretion spot on the WD and an extended stream or an extended magnetic accretion region on the WD surface. Such large eclipsing regions were also observed in other well known eclipsing polars like V895 Cen \citep{Stobie96}, V1432 Aql \citep{Watson95}, and HU Aqr \citep{Hakala93}.

The presence of double hump bright phase in the R-band light curve of RX J0859 suggests that both accreting poles are always visible and cyclotron radiation is present along the line of sight of the observer. The low flux observed during the phase $\sim$ 0.3-0.6 with respect to bright phases indicates that cyclotron emitting region might be behind the white dwarf during this time. The amplitudes of double hump bright phase are found to be changing from one observation to another indicating a variable accretion rate. The X-ray orbital modulations of RX J0859 also reveals two humps in an orbital cycle with different amplitude. The first hump was found to have a larger amplitude than the second hump. Double hump like structures have also been reported in the X-ray light curves of other polars MT Dra \citep{Schwarz02}, WX LMi \citep{Vogel07}, DP Leo \citep{Pandel02}, and  WW Hor \citep{Pandel02} etc., where the activity from the second pole is lower as compared to the primary in few polars, while few of them are equally accretes from both poles. Therefore, in the case of RX J0859, the unequal double hump in the light curve indicates that both poles of the WD emit the radiation unequally due to the different magnetic field strength and mass accretion rate.

R-band photometric light curves of RX J0749 show that the morphology of the eclipse profile changes from one epoch of observations to other. In the observation on 23 January 2015, the flat bottom eclipse profile was observed for RXJ0749 and where the width of the eclipse was observed to be $\sim$ 61 min. This duration is too long among polars to represent an eclipse of just the primary star. The value of $\Delta\phi_{ie}$ for RX J0749 is derived to be $0.16\pm0.01$. For an average value of q of 0.44 and $i \sim 80^{\circ}$, the value of R is estimated to be $\sim$ 50 R$_{WD}$ for RX J0749. As shown in Figure \ref{fig:rq}, for the extreme low and high values of q, R is estimated to be $\sim$ 88 R$_{WD}$ and  $\sim$ 23 R$_{WD}$, respectively for RX J0749.
 Thus the estimated radius of eclipsed light source represents the eclipse feature might be observed due to the obscuration of both an extended stream and hot compact accretion region. The fluctuations in the eclipse parameters like eclipse profiles, eclipse width, the duration of ingress and egress, and the dipper phasing are also observed in the light curves from all the observations. The brightness of RX J0749 is also decreased from the epoch 2014 to the epoch 2015, indicating that this system has undergone a transition from a high state in 2014 to a low state in 2015. These changes can tentatively be explained by the variation in the accretion rate, the size, density, and location of the accretion region. However, small changes in the stream trajectory will also produce changes in the absolute phasing of the dips. The photometric variations as seen on 23 January 2015 are also found to be similar to that of few other eclipsing polars like UZ For, MN Hya, V1309 Ori, and V1432 Aql etc. \citep[see][]{Bailey91, Buckley98, Staude01, Watson95}. Soft X-ray orbital modulation of RX J0749 exhibits a clear eclipse profile which seems to be closely resemble with R-band light curve. Though, both X-ray and optical observations are $\sim 17$ years apart, their eclipse profiles provide a hint that both emissions are originating from the same region.

The double-humped periodic orbital modulations are also observed in the optical light curves of RX J0649 and are more prominent than that of the RX J0859. As described above these double humped variations suggest that both accreted poles of WD emit the radiations at various wavelengths. However, both humps are almost equally strong in all observations of RX J0649 which suggests that both poles accrete at the same rate. The photometric light curves of RX J0649 are very similar to the eclipsing polar WW Hor \citep{Bailey88} and EP Dra \citep{Remillard91} where the light curves show a well-defined eclipse profile as well as two symmetric prominent maxima. Similar to WW Hor and EP Dra, the presence of two maxima in RX J0649 could also be consistent with that produced by beaming of cyclotron emission towards the direction of the observer which further confirms the identification of this object as polar. At first sight the light curves of RX J0649 also show an eclipse that look similar to an eclipse of a WD. However, for the derived value of  $\Delta \phi_{ie}$ of $0.15 \pm0.03$, the radius of the eclipsed source region is estimated to be $\sim$ 55 $R_{WD}$ with assumptions of $i \sim 80^{\circ}$ and $q$ = 0.54 for an average value of the mass of WD of 0.8 M$_\odot$ for RXJ0649. Figure \ref{fig:rq} also shows the variation of $R$ as a function of $q$ for RXJ0649. With  an extreme high value of $q$, $R$ is estimated to be more than 20 $R_{WD}$. This suggests that the hot accretion region is associated with the extended accretion stream and both are eclipsed by the secondary.  The soft X-ray light curve of  RXJ0649 was similar to its optical light curve, however, the eclipsed feature near the phase 0.0  in the X-ray light curve is shallower and narrow. Probably, this could be due to a smaller eclipsed region in the early epoch of observation.

The optical spectra of RX J0859, RX J0749, and RX J0649 are very similar to the spectra of other polars. They show strong emission lines of  Balmer series, He I,  and He II along with the conspicuous features of highly ionized optically thick regions like strong emission lines of He II 4686 \AA ~and CIII/NIII. Presence of these emission lines in the optical spectra  is a typical feature of AM Her systems \citep[e.g. see][]{Silber92}. The magnetic nature of these systems can be further confirmed by the presence of an inverse Balmer decrement, a large value of EW of the H$\beta$ ($>$ 20 \AA), and the ratio of He II(4686)/H$\beta$ $\textgreater$ 0.4 \citep[see][]{Silber92, Warner95}. Both He II 4686 \AA ~and H$\beta$ emission lines during a significant fraction of the orbit (except during the eclipse) also demonstrates that the emission lines are formed in  relatively compact regions which are illuminated by the strong photoionizing source like accretion column. The Balmer decrement, H$\beta$/H$\alpha$ was estimated in the range of 1.1-1.3 for these three sources which suggests that the lower limit of electron density of the order of 10$^{12}$ cm$^{-3}$ \citep[see][]{Cropper90}. These systems also appear to have been in relatively high states of accretion at the time of observations, as evidenced by the strength of their He II 4686 \AA ~emission line \citep{Patterson85}. Unlike Balmer emission lines, the Bowen fluorescence lines might be generated in these three sources due to the interaction of the X-rays with the surrounding gaseous matter in vicinity of the WD \citep[see][]{McClintock75}. The presence of the Bowen blend CIII/NIII together with the He II 4686 \AA ~emission line in each spectrum also suggests that the regions where the lines originate possess a high-temperature \citep{Schachter91}. The Fe II emission line is weak in the optical spectrum of RX J0749 and RX J0649, while almost absent in the spectrum of RX J0859. These Fe II emission lines could originate near the Roche lobe and be eclipsed by the secondary as the magnetic pole facing towards the observer. A similar behaviour has been observed in AM Herculis, where weak Fe II emission lines were  also detected in the out-of-eclipse phase \citep[see][]{Schmidt81}. 

Prominent broad emission humps are detected in the optical spectrum of each system. This could be due to the cyclotron emitting regions  which are viewed almost perpendicularly near the limb of the white dwarf \citep{Wickramasinghe91}. The identification of cyclotron humps in the optical spectra has revealed the magnetic field strength of about 49$\pm$2 MG, 43.5$\pm$1.4 MG, and 44$\pm$1 MG for RX J0859, RX J0749, and RX J0649, respectively, as is seen in many other polars. The cyclotron harmonics of RX J0859 and RX J0749 appear to closely resemble each other. In both systems, one cyclotron hump appears to be shallower than the other which could be observed due to the presence of the linear extension of the cyclotron emission region. The linearly extended cyclotron emission regions might be produced in the dispersed magnetic field, which could be responsible for the observed shallow feature in the optical spectrum of both RX J0859 and RX J0749 systems \citep{Wickramasinghe91}.  

\section{Conclusions}
We have determined the following properties of three poorly studied polars, namely RX J0859, RX J0749, and RX J0649, for the first time:
\begin{itemize}
\item All three X-ray sources are found to be eclipsing polars from their optical and X-ray observations.   
\item The orbital periods of RX J0859, RX J0749, and RX J0649 are 2.393$\pm$0.003 hrs, 3.672$\pm$0.001 hrs, and 4.347$\pm$0.001 hrs, respectively. Among these systems, RX J0859 is found to lie in the region of the period gap, while RX J0749 and RX J0649 are found to lie above the period gap.
\item The radius of the eclipsed light source is found to be more than the actual size of the WD in all the three polars, indicating that the eclipsed component is not only a  WD but also an extended accretion stream or an extended magnetic region on the WD surface.
\item The optical spectra of these systems are typical of polars, with strong high ionization emission lines and inverted Balmer decrement which confirms the magnetic nature of these systems.
\item The observed cyclotron humps in the optical spectrum, lead to an estimate of the magnetic field strength of 49$\pm$2 MG, 43.5$\pm$1.4 MG, and 44$\pm$1 MG for RX J0859, RX J0749, and RX J0649, respectively. 
\end{itemize}


\section{Acknowledgments}
We acknowledge the referee of this paper for his/her useful comments and suggestions. This research is based on the observations obtained by {\textit ROSAT}, which has been supported by the Bundesministerium f\"ur Foschung und Technolgie (BMFT) and the Max-Planck-Gesellschaft (MPG). One of the authors HPS thanks the Council of Scientific and Industrial Research (CSIR) for support.


\bibliographystyle{mnras}
\bibliography{ref}

\end{document}